%% file: Journal-vTComm.tex
\documentclass[journal,draftclsnofoot,onecolumn,12pt]{IEEEtran}
\usepackage{amsthm,amssymb,graphicx,multirow,amsmath,color,amsfonts}
\usepackage{setspace}
\usepackage[update,prepend]{epstopdf}
\usepackage{algorithm}

\usepackage[noend]{algpseudocode}
\usepackage[noadjust]{cite}
\usepackage[latin1]{inputenc}
\usepackage{tikz}
\usepackage{hyperref}
\usepackage{cleveref}

\usepackage{pdfpages}
\usepackage{tabulary}
\usepackage{multirow}
\usepackage{float}
\usepackage{filecontents}
\usepackage[font=footnotesize]{caption}
\usepackage{subcaption}
\usepackage{comment}
\usepackage[mathscr]{eucal}
\usepackage{mathtools}
\let\subparagraph\relax
\usepackage[compact]{titlesec}
\titlespacing{\section}{0pt}{*1}{*1}
\titlespacing{\subsection}{0pt}{*1}{*1}
\usepackage{setspace}

\include{notationITA}
\allowdisplaybreaks 
\title{{Tensor Learning-based Precoder Codebooks for FD-MIMO Systems}}

\author{
Keerthana Bhogi, Chiranjib Saha, and Harpreet S. Dhillon
\thanks{K. Bhogi and H. S. Dhillon are with Wireless@VT, Department of ECE, Virginia Tech, Blacksburg, VA, USA. C. Saha is with Technical Standards group, Qualcomm, San Diego, CA, USA.  Email: \{kbhogi, csaha, hdhillon\}@vt.edu. This work was performed when C. Saha was with Wireless@VT, Department of ECE, Virginia Tech, Blacksburg, VA, USA. The support of the US National Science Foundation (Grants ECCS-1731711 and CNS-1923807)  is gratefully acknowledged. A preliminary version of this paper was presented in Asilomar Conference on Signals, Systems and Computers, Pacific Grove, CA~\cite{bhogi2020learning}. 
} }

\begin{document}
\maketitle
\begin{abstract}
This paper develops an efficient procedure for designing low-complexity codebooks for precoding in a full-dimension (FD) multiple-input multiple-output (MIMO) system with a uniform planar array (UPA) antenna at the transmitter (Tx) using tensor learning. In particular, instead of using statistical channel models, we utilize a model-free data-driven approach with foundations in machine learning to generate codebooks that adapt to the surrounding propagation conditions. We use a tensor representation of the FD-MIMO channel and exploit its properties to design quantized version of the channel precoders. We find the best representation of the optimal precoder as a function of Kronecker Product (KP) of two low-dimensional precoders, respectively corresponding to the horizontal and vertical dimensions of the UPA, obtained from the tensor decomposition of the channel. We then quantize this precoder to design product codebooks such that an average loss in mutual information due to quantization of channel state information (CSI) is minimized. The key technical contribution lies in exploiting the constraints on the precoders to reduce the product codebook design problem to an unsupervised clustering problem on a Cartesian Product Grassmann manifold (CPM), where the cluster centroids form a finite-sized precoder codebook. This codebook can be found efficiently by running a $K$-means clustering on the CPM. With a suitable induced distance metric on the CPM, we show that the construction of product codebooks is equivalent to finding the optimal set of centroids on the factor manifolds corresponding to the horizontal and vertical dimensions. Simulation results are presented to demonstrate the capability of the proposed design criterion in learning the codebooks and the attractive performance of the designed codebooks. 
\end{abstract}


\section{Introduction}
With the availability of unprecendented amount of data, there is a significant interest in applying machine learning (ML) to a variety of problems in communications and signal processing~\cite{dorner2017deep,o2017introduction}. Many of these problems also have a rich history of research that has led to key insights about their general structures and properties, which are collectively referred to {\em domain knowledge}. It is well-acknowledged in the ML community that incorporating this domain knowledge in learning algorithms results in efficient solutions, which has generated significant interest around the general idea of {\em theory-guided ML}~\cite{karpatne2017theory}. The use of domain knowledge, such as the topological manifold on which the data is lying, often reduces the complexity of the ML models. 

In this paper, we explore the merger of domain knowledge and learning algorithm for the codebook design problem for limited feedback frequency division duplexing (FDD) MIMO systems. It is a classical problem in MIMO systems, where the CSI at the receiver (Rx) needs to be quantized before sending over the limited capacity feedback channel  to  the Tx  for  precoding~\cite{narula1998efficient}.  This  codebook  design  problem  has  been  studied  extensively under  several  statistical  channel  models  (see  \cite{love2008overview}  for  a  comprehensive  survey  on {\em model-based codebooks})  but  recently  gained  attention  from  the  perspective  of  ML.  The reason is that this problem can be viewed as a clustering problem where the set of optimal cluster centers represent the CSI whose distribution is available as a {\em training set}. Since the fundamental difficulty in this problem is the dimensionality of the channel, the natural tendency is to think in terms of obtaining a low dimensional representation of the channel using deep learning (DL) techniques, such as autoencoders, and use it for codebook construction~\cite{deepCSI2018, deepCSI2019}. An autoencoder operates on the hypothesis that the data possesses a representation on  a lower dimensional manifold (referred to as feature space), {\em albeit} unknown, and tries to learn the embedded manifold by training over the dataset~\cite[Chapter~14]{goodfellow2016deep}. In contrast, for MIMO beamforming and precoding, the underlying manifold is known to be a Grassmann manifold (GM) in some cases~\cite{bhogi2020learning,love2005limited}. This removes the requirement of ``learning'' the manifold from the dataset which often times can be extremely complicated. Once the manifold is known, we can leverage the ``shallow'' learning techniques like the clustering algorithms on the manifold to find the {\em precoder codebook}.

\subsection{Prior work}

In a limited feedback FDD-MIMO system, the assumption is that the Tx and Rx agree upon a common precoder codebook. The Rx, after the channel estimation, finds a precoder from this codebook and transmits the corresponding index over the feedback channel to Tx. There are various kinds of codebook design methods based on the above described two philosophies. 

{\em Model-based Approach. }For independently and identically distributed (i.i.d.) Rayleigh fading channels, the codebook design problem for precoding is equivalent to packing the subspaces in a GM of appropriate dimensions~\cite{love2003grassmannian, love2005limited}. For correlated channels, the Grassmann codebook can be modified by applying a channel correlation matrix~\cite{love2006limited, amiri2008adaptive}. The basis of this modification is the assumption that the channel matrix is assumed to be factored into the square-root channel correlation matrix (or the long-term statistics of the channel) and the i.i.d. Rayleigh fading channel (or the instantaneous CSI)~\cite{mcnamara2002spatial}. Apart from the Rayleigh fading assumption, another widely used channel model is the spatial channel model (SCM)~\cite{3gpp.25.996}, which has led to the design of discrete fourier transform (DFT) structured codebooks. The principle of DFT codebooks is to quantize the direction of arrival of the dominant radio path of the channel. Based on the same principle, more advanced hierarchical DFT codebooks were developed. One prominent example of hierarchical codebooks is the so-called double DFT codebooks, where the two codebooks are designed for quantizing the  long-term and instantaneous components of the precoder~\cite{shuang2011design}. While the codebooks were primarily developed for linear antenna arrays at Tx and Rx, for FD-MIMO systems these codebooks can be extended by the formulation of {\em product codebooks}. The product codebook is simply a product (such as KP) of two codebooks corresponding to the antenna arrays across the horizontal and vertical dimensions. The basis of this design is the Kronecker correlation model that approximates the channel correlation matrix with the KP of channel correlation matrices of horizontal and vertical dimensions. The decomposition of the channel correlation matrix  of UPA enables the natural extension of the existing codebooks, e.g. Grassmannian codebooks~\cite{ying2014kronecker} and  DFT   codebooks~\cite{li2013codebook, su2013limited,song2018advanced,choi2015advanced} for FD-MIMO systems.

{\em Data-driven Approach. }Unlike the model-based approach, a more direct approach for codebook design is to {\em learn} the codebooks from the channel datasets available through extensive channel measurements. The first comprehensive work in this direction is \cite{roh2006design}, where designing precoder codebooks is shown to be equivalent to a problem of vector quantization (VQ) on the space of optimal precoders i.e., right singular matrices of the channel matrices in the training dataset. In \cite{bhogi2020learning}, we have shown that this formulation has a natural connection to ML, since the codebook construction method is equivalent to Grassmannian $K$-means clustering~\cite{dhillon2003diametrical}. However, this technique is not useful when the number of antennas increases. This is because large number of antennas incur quantization or clustering in large dimensions which is not very efficient due to the {\em curse of dimensionality}~\cite{bellman2015adaptive}. As an alternate approach, the CSI compression has been cast as an autoencoder problem, where the encoder residing at the receiver compresses and quantizes CSI and decoder at Tx reconstructs the CSI. The extent of CSI compression of MIMO channels of arbitrary channel statistics and correlation properties in this scheme can be significantly enhanced by using deep neural network-based (more precisely, deep convolution neural networks (CNN)) structures for the encoder and decoder~\cite{deepCSI2018, deepCSI2019,yang2019deepCMC, guo2020convolutional}. Although these DL-based approaches have shown promising results compared to the state-of-the-art CSI compression techniques, their practical importance is questionable. The reason is that the performance is achieved only after using significantly complex architectures of the neural networks which is prone to a complicated hyperparameter tuning for any particular propagation environment.  While the CNN-based techniques were designed to operate on datasets which have natural interpretations in the Euclidean domain (such as images), we can extend CNNs to build autoencoders that operate on topological manifolds. However, it can be very challenging to design such models and still vastly considered as an open problem in ML. Therefore, in this paper, we propose an alternate formulation for the data-driven precoder design for FD-MIMO channels by building on the ideas of Grassmannian $K$-means clustering developed in the conference version~\cite{bhogi2020learning}. However, as we discussed before, extending this method for higher dimensions of channels is not straightforward. Interestingly, the FD-MIMO systems naturally admit a tensor representation of the channel~\cite{araujo2019tensor,de2003blind, STMCAlmeida}. This enables us to leverage tools from a more classical form of ML, known as {\em tensor learning}~\cite{TD2015Cichoki, sidiropoulos2017tensor, lui2012human}, along with ideas from theory-guided ML to constrain the outputs to a topological manifold, to formulate computationally efficient product codebooks for precoding even for large number of Tx antennas. 

\subsection {Contributions and Novelty}
In this paper, we propose a data-driven precoder codebook design method by exploiting a tensor representation of the FD-MIMO channel. We reduce the dimensionality of the channel tensor by decomposing it into low-dimensional orthonormal factors using the  low-rank Tucker decomposition (TD). This operation simplifies the codebook design explained as follows. 

First, the Rx computes the unquantized precoder from the channel tensor as a function of KP of the two low-rank TD factors corresponding to the horizontal and vertical dimensions of the UPA at the Tx. We adopt this KP structure of the unquantized precoders to the quantized precoders as well. We show that this KP structure of the precoders admits a representation on a {\em {Tensor Product Grassmann Manifold}} (TPM), where each factor is a GM corresponding to horizontal and vertical dimensions of the UPA at the Tx. We define a measure of loss in mutual information associated with an arbitrary precoder and use it to define the average mutual information loss due to the limited feedback, leading to a new codebook design criterion. With the rotational invariance property of the precoders and the induced chordal distance metric on a GM, we show that the obtained codebook design criterion is equivalent to minimizing the average distortion in representing the optimal unquantized precoders with quantized precoders on a TPM.

Second, we exploit the diffeomorphism between a TPM and a {\em {Cartesian Product Grassmann Manifold}} (CPM) to approximate the described quantization loss as the average distortion between the representations of the optimal unquantized and quantized precoder on the CPM. We show that the optimal product precoder codebook minimizing the defined average distortion due to quantization is equivalent to the set of optimal centroids given by the $K$-means clustering algorithm on the CPM. The induced chordal distance metric is inherited from the factor GMs to define the chordal distance on a CPM. This provides a natural extension of the $K$-means clustering algorithm on a GM to a CPM. With this induced chordal distance metric, we show that the $K$-means clustering problem on a CPM is reduced to separate $K$-means clustering problems on its factor manifolds. This simplifies the product precoder codebook construction to finding the optimal set of centroids using the $K$-means clustering on its factor manifolds corresponding to the horizontal and vertical dimensions of the UPA at the Tx. We also formally show that the proposed tensor based product codebook design is computationally more efficient than its VQ counterpart, proposed in~\cite{roh2006design}, in terms of asymptotic complexity. 

{\em Notations. } We use ${\bf a} \in {\C}^{M \times 1}$, ${\bf A} \in {\C}^{M \times N}$, to designate complex column vectors, matrices, respectively,  ${\bf A}(:,i)$ or ${\bf a}_i$ to denote the $i$-th column, ${\bf A}(:,i:j)$ to represent an $M \times (j-i+1)$ matrix,  formed by $i$-th to $j$-th columns of ${\bf A}$ for $1 \leq i \leq j \leq N$. If ${\cal I} = \{i_1,\cdots,i_n\}$ denotes a set of indices where $1 \leq i_1 < \cdots < i_n \leq N$, then ${\bf A}(:,{\cal I})$ or ${\bf A}_{\cal I}$ represents an \textbf{$M \times |{\cal I}|$} matrix  formed by the columns of ${\bf A}$ whose indices are given by ${\cal I}$. We use ${\cal U}(M,N)$, ${\cal U}_M$ to represent the set of all $M \times N$ complex orthonormal matrices, $M \times M$ unitary matrices, respectively. Further, $a^* ({\bf a}^*)$ denotes the complex conjugate of $a\in \C$ $({\bf a} \in {\C}^{M \times 1}$), ${\bf A}^{T}$, ${\bf A}^H$ denote transpose, Hermitian, ${\rm vec}({\bf A})$ denotes the vectorization of ${\bf A}$, $\E_A$ denotes expectation over the distribution of $A$ where $A$ is a random matrix or vector. Also, $|\cdot|, \norm{\cdot}_F$ denote the absolute value, the Frobenius norm and $j = \sqrt{-1}$.

\section{System Overview}

We consider a narrow-band point-to-point MIMO communication system, where the Tx and Rx are equipped with $M_t$ and $M_r$ antennas, respectively. We assume a block fading channel model and represent the channel between Tx and Rx as ${\bf H} \in \C^{M_r \times M_t}$. Throughout this paper, we assume that $M_r \leq M_t$ and let the rank of the channel matrix ${\bf H}$ be $\texttt{r}_o \leq M_r$. The Tx is equipped with a UPA antenna with $M_v$ and $M_h$ antennas in the vertical and horizontal dimensions, respectively with $M_t = M_vM_h$ and the Rx is equipped with a ULA antenna with $M_r$ antennas. The discrete-time baseband input-output relation for this system can be expressed as ${\bf y} = {\bf H}{\bf x} + {\bf n}$, where ${\bf x}\in \C^{M_t \times 1}$ is the transmitted signal, ${\bf y}\in \C^{M_r \times 1}$ is the received signal and ${\bf n}\in \C^{M_r \times 1}$ is the additive white Gaussian noise distributed as ${\cal CN}({\bf 0},N_o{\bf I}_{M_r})$. The average total transmit power is denoted as ${\cal E}_s$ where ${\cal E}_s = \E[{\bf x}^H{\bf x}]$. The SVD of ${\bf H}$ is given by
    ${\bf H} = {\bf U}{\bf \Sigma}{\bf V}^H,$  
where ${\bf U} \in {\cal U}_{M_r}$, ${\bf V} \in {\cal U}_{M_t}$, and ${\bf \Sigma}$ is the $M_r \times M_t$ rectangular diagonal matrix with $i$-th largest singular value $\sigma_i$ at the entry ${(i, i)}$.

\subsection{Beamforming}\label{sec::BF}
For the simplicity of exposition, we first consider a multiple-input single-output (MISO) system, where the Rx is equipped with a single antenna i.e., $M_r = 1$. In order to improve the received SNR, the Tx performs beamforming. For this case, the received signal ${\bf y}$ simplifies to ${y} = {\bf H}{\bf f}s + {n}$, where $s \in {\C}$ is the transmitted symbol with average power $\E_{s}[s^*s] = {\cal E}_s$, ${\bf f} \in {\C}^{M_t \times 1}$ is the beamformer. Assuming that the Rx employs maximal ratio combining (MRC)~\cite{love2003grassmannian}, the Rx uses ${z} = \frac{{\bf H}{\bf f}}{\norm{{\bf H}{\bf f}}_2}$ to estimate the transmitted symbol $\hat{s}$ which is simplified as $\hat{s} = {z}^H{y} = {y}$. This gives the receive SNR $\rho_r$ as $\rho_r =  {\cal E}_s\frac{\norm{{\bf H}{\bf f}}^2_2}{\norm{{\bf n}}^2_2\norm{{\bf f}}^2_2} = \frac{{\cal E}_s}{N_o} \frac{\norm{{\bf H}{\bf f}}^2_2}{\norm{{\bf f}}^2_2} = \rho_t \frac{\norm{{\bf H}{\bf f}}^2_2}{\norm{{\bf f}}^2_2}$ where ${\cal E}_s/N_o$ is the transmit SNR $\rho_t$. The total transmit power $\E[{\bf x}^H{\bf x}] = \E[{\norm{{\bf f}{s}}^2_2}] = {\cal E}_s$ is assumed to be fixed. Because of this, we have the unit norm constraint on the beamformer, i.e., $\norm{{\bf f}}^2_2 = 1$ and thus ${\bf f} \in {\cal U}(M_t, 1)$. Following this constraint, the beamforming gain ${\Gamma}({\bf H}, {\bf f})$ is obtained as ${\Gamma}({\bf H}, {\bf f}) := {\rho_r}/ {\rho_t} = \norm{{\bf H}{\bf f}}^2_2$. The problem of transmit beamforming is to  maximize ${\Gamma}({\bf H}, {\bf f})$ i.e., $\hat{\bf f} = \underset{{\bf f} \in {\cal U}(M_t, 1)} \argmax\ {\Gamma}({\bf H}, {\bf f}) = \underset{{\bf f} \in {\cal U}(M_t, 1)} \argmax\ \norm{{\bf H}{\bf f}}^2_2$. One possible solution for the optimal beamformer $\hat{\bf f}$ is the right singular vector that is associated with the maximum singular value of ${\bf H}$ i.e., $\hat{\bf f} = {\bf v}_1 = {\bf V}(:,1)$~\cite{tse2000performance}. The corresponding beamforming gain is ${\Gamma}_{\rm max} = \underset{{\bf f} \in {\cal U}(M_t, 1)} \max\ {\Gamma}({\bf H}, {\bf f}) = {\Gamma}({\bf H}, {\bf v}_1) = \norm{{\bf H}{\bf v}_1}_2^2 = \sigma^2_{1}$. For transmit beamforming, it has been shown that the beamformer that maximizes the receive SNR $\rho_r$ also maximizes the mutual information between $s$ and $y$ and minimizes the average probability of symbol error~\cite{andersen2000antenna,simon2005digital}. 

\subsection{Precoding}\label{sec::Precoding}
Let us now consider a general MIMO system with $M_r > 1$. Since $M_r > 1$, the system can support upto ${\rm rank}-\texttt{r}\ (1 \leq \texttt{r} \leq M_r)$ transmission or the transmission of $\texttt{r}$ independent streams. For this scheme, we assume transmit precoding, i.e., the Tx transmits ${\bf s} \in {\C}^{\texttt{r} \times 1}$, a symbol vector of $\texttt{r}$ independent data streams, which is precoded with a precoder matrix ${\bf F}\in {\C}^{M_t \times \texttt{r}}$. The transmitted signal ${\bf x}$ is obtained as ${\bf x} = {\bf F}{\bf s}$ resulting in the received signal ${\bf y} = {\bf H}{\bf F}{\bf s} + {\bf n}$. We assume equal power allocation strategy at the Tx where the total transmit power ${\cal E}_s$ is split equally among the $\texttt{r}$ transmitted symbols i.e., $\E_{s_i}[s^*_is_i] = \frac{{\cal E}_s}{\texttt{r}}$ and also assume that ${\bf s}$ is generated by an uncorrelated zero-mean jointly Gaussian symbol source. Thus, ${\bf s} \sim {\cal N}({\bf 0}, {\frac{{\cal E}_s}{\texttt{r}}}{\bf I}_{\texttt{r}})$. When the Tx precodes ${\bf s}$ with ${\bf F}$, the equivalent channel is ${\bf H}_{eq} = {\bf H}{\bf F}$ and the transmit SNR per spatial stream is $\rho_t = \frac{{\cal E}_s}{N_o\texttt{r}}$. The Rx uses a linear minimum mean square error (MMSE) combiner to estimate the transmitted symbol vector as $\hat{\bf s} = {\bf Z}^H_{\rm MMSE}{\bf y}$ where ${\bf Z}_{\rm MMSE} = {\bf H}^H_{eq}\left({\bf H}^H_{eq}{\bf H}_{eq} + {\rho_t}^{-1}{\bf I}\right)^{-1}$.  Under these assumptions, the mutual information $R({\bf H},{\bf F})$ between ${\bf s}$ and ${\bf y}$ for a given channel ${\bf H}$ and a precoder ${\bf F}$ is given by 
\begin{align}
    R({\bf H},{\bf F}) &= \log{\rm det}\left({\bf I} + \rho_t{\bf H}_{eq}^H{\bf H}_{eq}\right) = \log{\rm det}\left({\bf I} + \rho_t{\bf F}^H{\bf H}^H{\bf H}{\bf F}\right). \notag 
\end{align}
With full CSI at the Tx (CSIT), the strategy that maximizes the mutual information $R({\bf H}, {\bf F})$ is to employ water-filling based optimal power allocation on the $\texttt{r}$ independent data streams~\cite{cover1999elements,scaglione2002optimal}. This necessitates the knowledge of $\bar{\bf V} = {\bf V}(:, 1:\texttt{r})$ and additionally ${\bf \Sigma}$, truncated upto $\texttt{r}$ dominant singular values, to ensure optimal power splitting across the spatial streams at the Tx for precoding. 

\begin{figure}
\centering
\begin{minipage}{.48\textwidth}
  \centering
    \includegraphics[width = 0.8\textwidth]{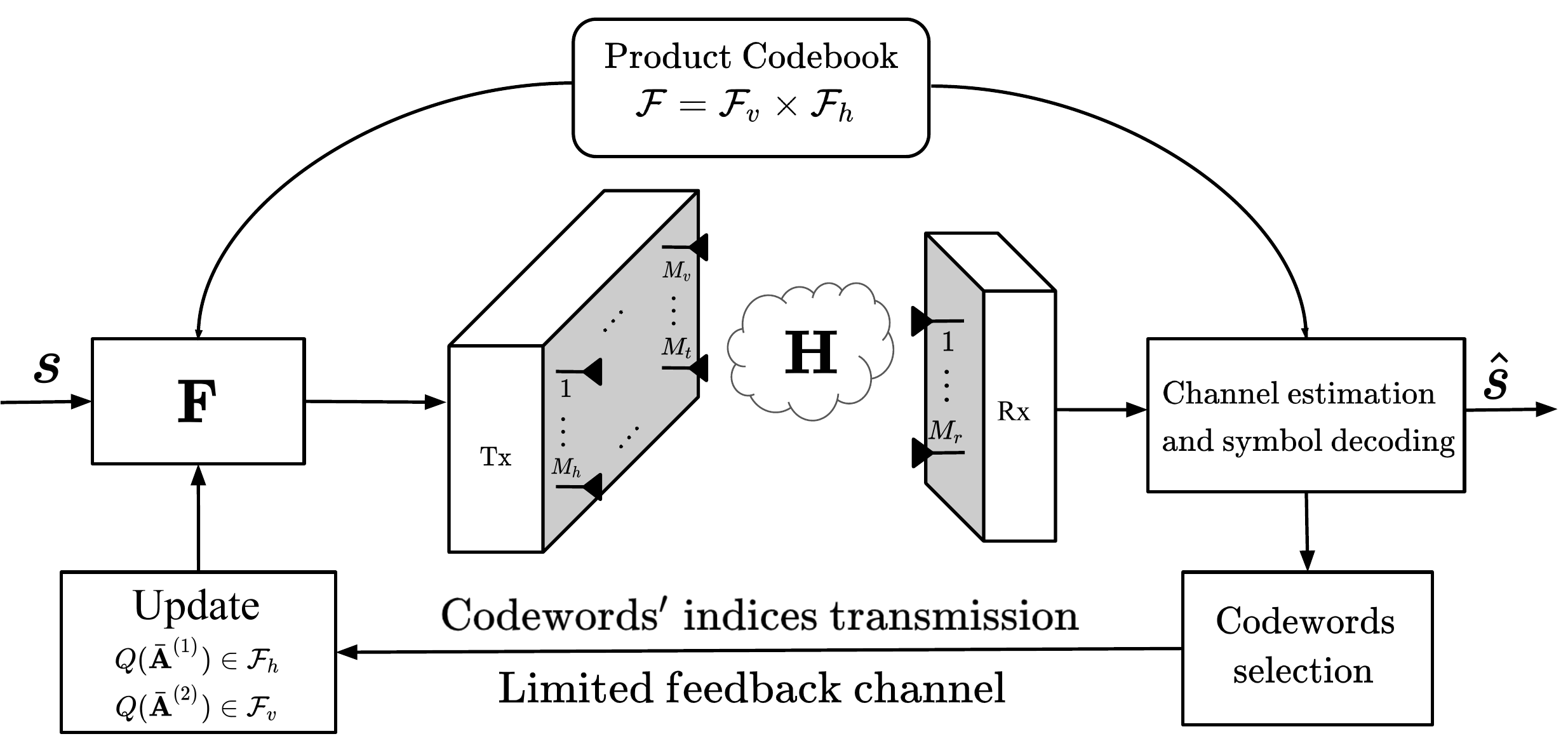}
    \caption{Block diagram of an FDD-MIMO system with limited feedback channel of capacity $B$ bits per channel use.}
    \label{fig::BlockDiagram}
\end{minipage}%
\hfill
\begin{minipage}{.48\textwidth}
  \centering
    \includegraphics[width = 0.8\textwidth]{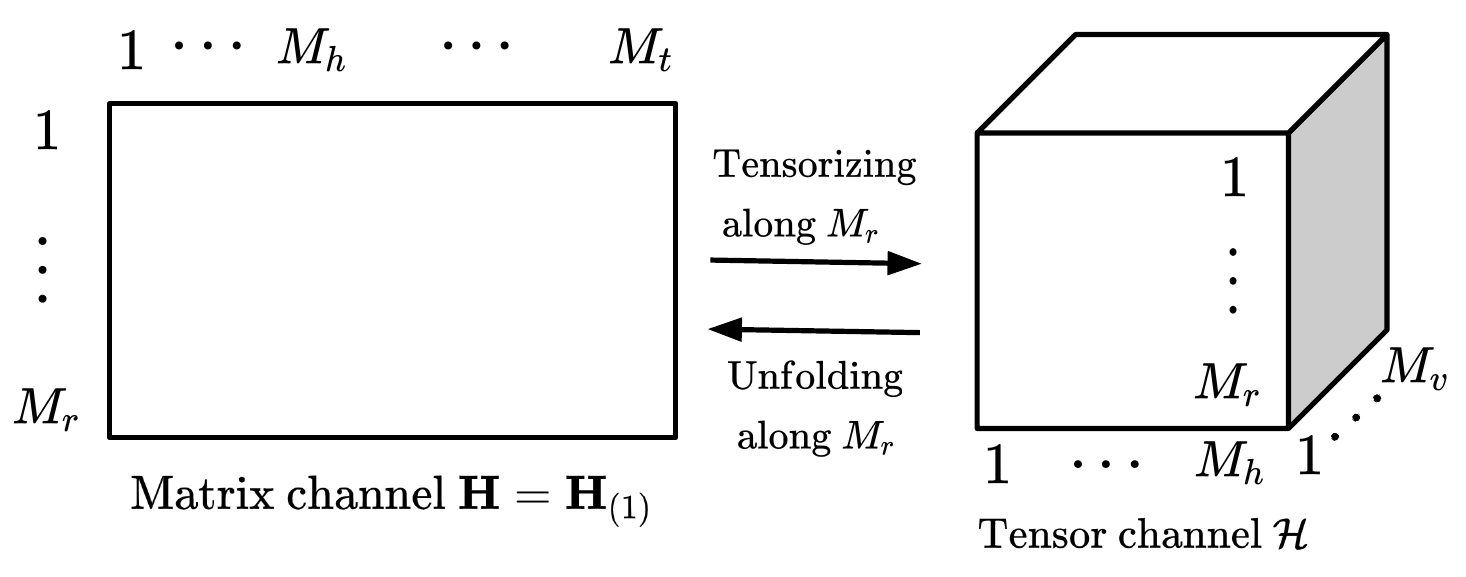}
    \caption{Tensor representation of FD-MIMO channel}
    \label{fig::Tensor::Channel}
\end{minipage}
\end{figure}

For the optimal beamforming (precoding), the Tx needs to know ${\bf v}_1$ $(\bar{\bf V}, \bar{\bf \Sigma})$. In an FDD system, the Rx estimates the channel ${\bf H}$ and sends ${\bf v}_1$ $(\bar{\bf V}, \bar{\bf \Sigma})$ back to the Tx over a feedback channel. Thus the feedback overhead increases as $M_t$ increases. Since the feedback channel is typically assumed to be a low-rate, zero-delay, and error-free, with a limited capacity of $B$ bits per channel use, it is not always possible to transmit ${\bf v}_1$ $(\bar{\bf V}, \bar{\bf \Sigma})$ over this channel without any data compression, especially when the number of antennas is large~\cite{narula1998efficient}. Thus, it is necessary to introduce some method to quantize ${\bf v}_1$ $(\bar{\bf V}, \bar{\bf \Sigma})$. The available $B$ feedback bits per each channel use have to be utilized to convey the channel information to the Tx and maximize the performance of the MIMO system. The most well-known approach for the quantization is to construct a finite-sized dictionary of beamformers (precoders) \cite{narula1998efficient}, also known as the {\em codebook}. In particular, for beamforming, the Tx and Rx agree upon a beamformer codebook, say ${\cal F} = \{{\bf f}_1, \dots ,{\bf f}_{2^B}\}, {\bf f}_i \in {\cal U}(M_t, 1)$. While there are multiple ways to define a precoder codebook for quantizing $\bar{\bf V}$, we focus on the most common approach of orthonormal precoder codebook where the precoders are always constrained to be orthonormal matrices\footnote{With limited feedback bits available, we focus first on representing $\bar{\bf V}$ and do not allocate any bits for power allocation information i.e., $\bar{\bf \Sigma}$, thus assuming equal power allocation strategy.}~\cite{kapetanovic2010comparison,love2005limited}. The orthonormality constraint follows from the form of the optimal precoders derived with the maximum eigenvalue constraint on ${\bf F}$ under the presence of full CSIT~\cite{scaglione2002optimal}. Under the equal power allocation strategy and the orthonormality contraints on ${\bf F}$, an optimal ${\rm rank}-\texttt{r}$ precoder over ${\cal U}(M_t, \texttt{r})$ that maximizes the mutual information $R({\bf H}, {\bf F})$ is ${\bf F}_{\rm opt} = \bar{\bf V}$ which is formed by the $\texttt{r}$ dominant columns of ${\bf V}$~\cite{scaglione2002optimal}. Thus a codebook ${\cal F}$ of cardinality $2^B$ with candidate precoder matrices is given as ${\cal F} = \{{\bf F}_1, \dots ,{\bf F}_{2^B}\}$, where ${\bf F}_i \in {\cal U}(M_t, \texttt{r})$  and is assumed to be known to the Tx and Rx. The Rx chooses the appropriate beamformer ${\bf f} \in {\cal F}$ $\big($precoder ${\bf F} \in {\cal F}\big)$ that maximizes $\Gamma({\bf H}, {\bf f}) \left(R({\bf H}, {\bf F})\right)$ and feeds the index of the codeword back to the Tx.  For a given beamformer codebook ${\cal F}$, the criterion for choosing the optimal beamformer can be stated as ${\bf f} = \underset{{\bf f}_i \in {\cal F}} \argmax\ {\Gamma}({\bf H}, {\bf f}_i) = \underset{{\bf f}_i \in {\cal F}} \argmax\ \norm{{\bf H}{\bf f}_i}^2_2$. Similarly, for a given precoder codebook ${\cal F}$, the criterion for choosing the optimal precoder is ${\bf F} = \underset{{\bf F}_i \in {\cal F}} \argmax\ R({\bf H}, {\bf F}_i)$. The system-level diagram of a limited feedback FDD-MIMO system is provided in Fig.~\ref{fig::BlockDiagram}.

\section{Preliminaries}
In this section, we briefly review the background of the topics including a few useful results that are used in developing the codebook design scheme proposed in the sequel. 

\subsection{Tensors}\label{sec::Tensors}
A tensor is a multi-dimensional array and the number of dimensions of the array is defined as the order of the tensor. A matrix, for instance is a two-dimensional array or second-order tensor. We denote an $N$-th order tensor complex tensor as ${\nscrX} \in {\C}^{I_1 \times \cdots \times I_n \times \cdots I_N}$ whose $(i_1,\cdots,i_n, \cdots,i_N)$-th element is represented as $x_{i_1 i_2 \cdots i_N}$ or $[\nscrX]_{i_1 i_2 \cdots i_N}$, where $1 \leq i_n \leq I_n$ for $n = (1,\cdots,N)$. The Frobenius norm of a tensor ${\nscrX}$ is denoted as $\norm{\nscrX}_F$ and defined as the square root of the sum of the squares of absolute values of its elements i.e., $\norm{\nscrX}_F := \sqrt{\sum^{I_1}_{i_1 = 1}\cdots\sum^{I_N}_{i_N = N}|x_{i_1\cdots i_N}|^2}$.

 A tensor can be represented by a set of matrices which is possible through unfolding the tensor. The rows and columns of a matrix are generalized as mode-$n$ fibers of a tensor. A mode-$n$ fiber is formed by the set of elements of the tensor where $i_n = (1,\cdots,I_n)$ for a chosen $i_1,\cdots,i_{n-1},i_{n+1},\cdots,i_N$. The unfolding of a tensor $\nscrX$ along its $n$-th dimension is called mode-$n$ unfolding and the resultant matrix is denoted as ${\bf X}_{(n)} \in {\C}^{I_n \times J_n}$ where $J_n = \prod \limits^{N}_{k=1,k\neq n} I_k$. The matrix ${\bf X}_{(n)}$ is formed by arranging the mode-$n$ fibers of $\nscrX$ as its columns. An element $x_{i_1 i_2\cdots  i_N}$ of ${\nscrX}$ is mapped to ${(i_n, j)}$-th element of ${\bf X}_{(n)}$ where $j = 1 +\sum^{N}_{k = 1,k \neq n}(i_k-1)J_k, J_k = \prod^{k-1}_{m=1,m \neq n}I_m$. The product of a tensor and a matrix along the $n$-th dimension is represented as ${\times_n}$ and known as $n$-mode product. The $n$-mode product of a tensor ${\cal X}$ and a matrix ${\bf U} \in {\mathbb C}^{J \times I_n}$ is represented as $ {\nscrY} = {\nscrX} \times_n {\bf U}$ where $\nscrY \in {\C}^{I_1 \times \cdots I_{n-1} \times J \times I_{n+1} \times \cdots I_N}$ whose mode-$n$ unfolding is given by ${\bf Y}_{(n)} = {\bf U}{\bf X}_{(n)}$.

{\em Tucker decomposition of a tensor. }TD decomposes a tensor into a core tensor and a set of orthonormal matrices corresponding to each mode of the tensor. It is also a form of higher-order principal component analysis~\cite{kapteyn1986approach} and TD of a tensor ${\nscrX}$ is expressed as ${\nscrX} = {\nscrG} \times_1 {\bf A}^{(1)} \times_2 {\bf A}^{(2)}\cdots \times_{N} {\bf A}^{(N)}$, for $i_n = (1,\cdots,I_n)$, $n=(1,\cdots,N)$. The tensor ${\nscrG} \in {\C}^{I_1 \times \cdots I_n \times \cdots I_N}$ is called the {\em core tensor} and the factor matrices ${\bf A}^{(n)} \in {\cal U}_{I_{n}}$. Let ${\bf G}_{(n)}$ be the mode-$n$ unfolding of ${\nscrG}$, then, from the TD of ${\nscrX}$ we have, ${\bf X}_{(n)} = {\bf A}^{(n)}{\bf G}_{(n)}\left({\bf A}^{(N)} \otimes \cdots \otimes {\bf A}^{(n+1)}\otimes{\bf A}^{(n-1)} \otimes \cdots \otimes {\bf A}^{(1)}\right)^{T}$. The matrices ${\bf A}^{(n)}$ can be thought of as the {\em principal components} in each mode and are analogous to principal components of a matrix. The core tensor ${\nscrG}$ represents the interaction between different principal components of $\nscrX$ and generally not a diagonal matrix as it is in the SVD of matrices.

{\em Low-rank representation. } A tensor ${\nscrX} \in {\C}^{I_1 \times \cdots \times I_N}$ can be approximated with a ${\rm rank}-(r_1,\cdots,r_N)$ tensor $\bar{\nscrX}$ as ${\nscrX} \approx \bar{\nscrX} = \bar{\nscrG} \times_1 {\bf A}^{(1)}_{r_1} \times_2 {\bf A}^{(2)}_{r_2} \cdots \times_{N} {\bf A}^{(N)}_{r_N}$ where $\bar{\nscrG} \in {\C}^{r_1 \times \cdots \times r_n \times \cdots \times r_N}$, $r_{n} \leq I_{n}$ for $n = (1,\cdots,N)$ and ${\bf A}^{(n)}_{r_n} \in {\cal U}(I_n,r_n)$ is a ${\rm rank}-r_n$ orthonormal matrix. The best ${\rm rank}-(r_1,\cdots,r_N)$ approximation ${\bar{\nscrX}}$ of ${\nscrX}$ is obtained as
\begin{align}
    (\bar{\nscrG}, {\bf A}^{(1)}_{r_1}, \cdots, {\bf A}^{(N)}_{r_N}) &= \underset{\bar{\nscrG},{\bf A}^{(i)}_{r_i} \in {\cal U}(I_i,r_i)}\argmin \norm{{\nscrX} - \bar{\nscrG} \times_1 {\bf A}^{(1)}_{r_1} \times_2 {\bf A}^{(2)}_{r_2} \cdots \times_{N} {\bf A}^{(N)}_{r_N}}_F. \label{eq::TDecomp_LR::Opt}
\end{align}
In the case of matrices, the principal components of the best low-rank approximation are obtained directly from its SVD~\cite{eckart1936approximation}, whereas for tensors, the above minimization problem has to be solved for obtaining the principal components of the tensor. One of the algorithms utilized for solving \eqref{eq::TDecomp_LR::Opt} is the {\em {Higher-Order Orthogonal Iteration}} (HOOI), which will be used in the sequel~\cite{kolda2009tensor}. 

\subsection{Overview of Grassmann Manifolds}\label{sec::GrassmannManifolds}
The {\em {complex}} GM ${\cal G}(n, k)$~\cite{love2005limited} is defined as the set of all $k$ dimensional linear subspaces spanned by orthonormal matrices ${\cal U}(n, k)$ i.e., ${\cal G}(n, k) := \{{\rm span}({\bf F}): {\bf F} \in {\cal U}(n, k) \}$, where ${\rm span}({\bf F})$ is the $k$ dimensional subspace in ${\C}^{n}$ spanned by the columns of the orthonormal basis ${\bf F}$. For any ${\bf Q} \in {\cal U}_k$, ${\rm span}({\bf F}{\bf Q}) = {\rm span}({\bf F})$, i.e., the subspaces spanned by the columns of ${\bf F}$ and ${\bf F}{\bf Q}$ are the same and are represented by an equivalence relation ${\bf F} \sim {\bf F}{\bf Q}$. Therefore the matrix representation of a point in ${\cal G}(n, k)$ is not unique. We use the notation ${\bf F} \in {\cal G}(n, k)$ to represent the subspace ${\rm span}({\bf F})$. Let ${\bf F}_1, {\bf F}_2 \in {\cal G}(n, k)$, then the distance between the subspaces spanned by them is characterized by the principal angles between ${\rm span}({\bf F}_1), {\rm span}({\bf F}_2)$. A number of different geodesic distances between the subspaces can be defined. In this paper, we will be using the {\em {chordal distance}}. The chordal distance $(d_c)$ between two subspaces which are spanned by ${\bf F}_1, {\bf F}_2 \in {\cal U}(n, k)$ is defined as $d^2_c({\bf F}_1, {\bf F}_2) := \frac{1}{2}\norm{{\bf F}_1{\bf F}^H_1 - {\bf F}_2{\bf F}^H_2}^2_F = \left(k - \norm{{\bf F}^H_1{\bf F}_2}^2_F\right) = \norm{\sin{\bf {\Theta}}}^2_2$, where ${\bf {\Theta}} = [\theta_1,\cdots,\theta_k]$ and $\theta_i$ is the $i$-th principal angle between ${\rm span}({\bf F}_1)$ and ${\rm span}({\bf F}_2)$. Any element on a GM is invariant to rotations i.e., ${\bf F} \equiv {\bf F}{\bf Q}$ for ${\bf Q} \in {\cal U}_k$. Therefore the chordal distance $d_c({\bf F}_1, {\bf F}_2)$ is invariant under various representations of the subspaces, i.e., $d_c({\bf F}_1, {\bf F}_2) = d_c({\bf F}_1{\bf Q}_1, {\bf F}_2{\bf Q}_2)$ $\forall$ ${\bf Q}_1, {\bf Q}_2 \in {\cal U}_k$.

\subsubsection{Product Grassmann Manifolds}
The $m$-fold CPM ${\cal G}^{\times}({\bf n}, {\bf k})$ is defined as the space  ${\cal G}(n_1, k_1) \times \cdots \times {\cal G}(n_m, k_m)$. A point in ${\cal G}^{\times}({\bf n}, {\bf k})$ is represented as the collection of the points ${\bf F}_i \in {\cal G}(n_i, k_i)\ \forall i = 1,\cdots,m$. Thus,
\begin{align}
    {\cal G}^{\times}({\bf n}, {\bf k}) &:= \{[{\bf F}] = ({\bf F}_1,\cdots, {\bf F}_m) | {\bf F}_i \in {\cal G}(n_i, k_i), i = 1,\cdots,m\}, \label{eq::directproduct::grassmann::def}
\end{align}
where $({\bf n}, {\bf k}) := \left((n_1, k_1), (n_2,k_2), \cdots, (n_m,k_m)\right)$. Just as different notions of distances on a GM~\cite{edelman1998geometry}, a distance metric on a CPM can be defined in different ways. We extend the chordal distance metric $d_c$ on a GM to define the following distance metric to measure the distance between two points $[{\bf F}]$, $[{\bf F}'] \in {\cal G}^{\times}({\bf n}, {\bf k})$: $
    d_c([{\bf F}], [{\bf F}']) := \norm{\sin{\bf {\Theta}}}_2,$ 
where ${\bf {\Theta}} =({\boldsymbol{\theta}}_1,\cdots,{\boldsymbol{\theta}}_m)$, ${\boldsymbol{\theta}}_i$ is the set of principal angles between the $i$-th factor GM of $[{\bf F}]$ and $[{\bf F}']$, i.e., ${\bf F}_i$ and ${\bf F}'_i$ respectively. Using this expression, the chordal distance on a CPM can also be written as
\begin{align}
    d^2_c([{\bf F}], [{\bf F}']) &= d^2_c\left(({\bf F}_1,\cdots, {\bf F}_m), ({\bf F}'_1,\cdots, {\bf F}'_m)\right) =  \sum^{m}_{i = 1}d^2_c({\bf F}_i, {\bf F}_i'). \label{eq::ChordalDistance::DPGM}
\end{align}

It implies that the squared chordal distance between two points on a CPM is equivalent to the sum of squares of distance between the points on the factor GMs that form the product space. This property will be particularly useful in the proposed product codebook construction. In the sequel, we will introduce another type of product GM, termed TPM, while designing the product codebook. 

\subsubsection{$K$-means Clustering on a Grassmann Manifold}\label{sec::Kmeans}
The $K$-means clustering on a given metric space is a method of VQ to partition a set of $N$ data points into $K$ non-overlapping clusters, in which each data point belongs to the cluster with the nearest cluster centroid. The centroids are the quantized representations of the data points that belong to the respective clusters. A quantizer on the given metric space maps the data points to one of the $K$ centroids. The $K$ centroids are chosen such that the average distortion due to quantization is minimized. Before we formally introduce the main steps of the clustering algorithm on ${\cal G}(n, k)$, we first define the notion of a distortion measure and a quantizer as follows.

\begin{ndef}[Distortion measure]\label{def::distortion::function}
The distortion caused by representing ${\bf F} \in {\cal G}(n, k)$ with ${\bf F}'\in {\cal G}(n, k)$ is defined as the distortion measure $d_o$ which is given by $d_o({\bf F}, {\bf F}') = d_c^2({\bf F}, {\bf F}')$.
\end{ndef}

\begin{ndef}[Grassmann quantizer]\label{def::CSI::quantizer}
Let ${\cal F} \subseteq {\cal G}(n, k)$ be a $B$-bit codebook such that ${\cal F} = \{{\bf F}_1,....,{\bf F}_{2^B}\}$, then a Grassmann quantizer $Q_{\cal F}$ is defined as a function mapping elements of ${\cal G}(n, k)$ to elements of ${\cal F}$ i.e., $Q_{\cal F}: {\cal G}(n, k) \mapsto {\cal F}$.
\end{ndef}
A performance measure of a Grassmann quantizer is the average distortion $D(Q_{\cal F})$, where $D(Q_{\cal F}) := \E_{{\bf X}}[d_o({\bf X}, Q_{\cal F}({\bf X})] = \E_{{\bf X}}[d^2_c({\bf X}, Q_{\cal F}({\bf X})]$. In most practical settings, we may have access to a set of $N$ data points ${\cal X} = \{{\bf X}\} \subseteq {\cal G}(n, k)$ in lieu of the probability distribution $p({\bf X})$. Then the expectation w.r.t ${\bf X}$ in $D(Q_{\cal F})$ means averaging over the set ${\cal X}$. Therefore the objective of $K$-means clustering with $K = 2^B$ is to find the set of $K$ centroids, i.e., ${\cal F}^{K}$, that minimizes $D(Q_{\cal F})$ and can be expressed as
\begin{align}
\scalebox{0.9}{$
    {\cal F}^{K} =  \underset{{\cal F}\subseteq {\cal G}(n, k)|{\cal F}| = 2^B} \argmin\ D(Q_{\cal F}) = \underset{{\cal F}\subseteq {\cal G}(n, k)
     |{\cal F}| = 2^B} \argmin\ \E_{{\bf X}} \big[ d^2({\bf X}, Q_{\cal F}({\bf X})) \big]$}, \label{eq::Kmeans::objective}
\end{align}
and the associated quantizer is $Q_{{\cal F}^{K}}({\bf X}) = \underset{{\bf F}_i \in {\cal F}} \argmin\ d_o({\bf X}, {\bf F}_i)  = \underset{{\bf F} \in {\cal F}} \argmin\ d_c^2({\bf X},{\bf F}_i)$. However, finding the optimal solution for $K$-means clustering is an NP-hard problem. Therefore, we use the Linde-Buzo-Gray algorithm \cite{LBGAlgorithm} (outlined in Alg.~\ref{alg::Product::KmeansAlgo}) which is a heuristic algorithm that iterates between updating the cluster centroids and mapping a data point to the corresponding centroid that guarantees convergence to a local optimum. In Alg.~\ref{alg::Product::KmeansAlgo}, the only non-trivial step is the centroid calculation for a set of points. In contrast to the squared distortion measure in the Euclidean domain, the centroid of a set of elements in a general manifold with respect to an arbitrary distortion measure does not necessarily exist in a closed form. However, the centroid computation on ${\cal G}(n, k)$ is feasible because of the following lemma~\cite{mondal2007quantization}.

\begin{lemma}[Centroid computation]\label{lem::Centroid::Computation} For a set of points ${\cal S}_i = \{{{\bf X}_j}\}^{N_k}_{j=1}$, ${\bf X}_j \in {\cal G}(n, k)$, that form the $i$-th Voronoi partition, the centroid ${\bf F}_i$ is ${\bf F}_i = \underset{{\bf F} \in {\cal G}(n, k)} \argmin \sum^{N_k}_{j=1} d_c^2({\bf X}_j,{\bf F}) = {\rm eig}_r\left(\sum^{N_k}_{j=1} {\bf X}_j{\bf X}^H_j \right),$
where the columns of ${\rm eig}_r({\bf Y})$ are chosen to be the $r$ dominant eigenvectors of the ${\bf Y}$.
\end{lemma}

\subsection{Submodular Optimization}\label{sec::Submodularity}
We now introduce a special form of optimization of {\em {set functions}} which will be a necessary building block of our proposed codebook design scheme. Consider a set function $f: 2^{\cal V} \mapsto \R$ which assign a real value to any subset ${\cal P}$ of a finite {\em {ground set}} ${\cal V} \neq \emptyset$. Then a function $f$ is called {\em monotone} if $f({\cal P} \cup \{a\}) - f({\cal P}) \geq 0$ for all ${\cal P} \subseteq {\cal U}$, $a \notin {\cal P}$ and $a \in {\cal V}$. Further, a set function $f$ is {\em submodular} if $f({\cal P} \cup \{a\}) - f({\cal P}) \geq f({\cal T} \cup \{a\}) - f({\cal T})$ for all possible pairs of subsets ${\cal P} \subseteq {\cal T} \subseteq {\cal V}$ and all elements $a \in {\cal V}$, $a \notin {\cal T}$. Intuitively, submodularity refers to the law of diminishing return: the marginal gain of $f({\cal P})$ by adding an element $a$ to $\cal P$ diminishes as the size of $\cal P$ increases for all $P$. The submodular maximization problem subjected to the cardinality constraint can be formulated as follows: ${\cal P}^* = \underset{{\cal P} \subseteq {\cal U}, |{\cal P}| = n} \argmax\ f({\cal P})$. Submodular optimization problems are known to be NP-hard~\cite{nemhauser1978best}. However, there exist greedy algorithms with a linear complexity ${\cal O}\left(|{\cal U}||{\cal P}|\right)$~\cite{nemhauser1978analysis}, which  achieve atleast a $(1-1/e)$-factor approximation of the optimal solution.

\section{Product Codebook Design for Beamforming}
To enable the CSIT for beamforming (precoding) through codebooks, a quantization scheme for quantizing the optimal beamformer (precoder) and a design criterion for constructing the respective codebooks are necessary. An efficient iterative beamformer (precoder) codebook design method based on vector quantization of the space ${\C}^{M_t \times 1}$ $({\C}^{M_t \times \texttt{r}})$ is proposed in~\cite{roh2006transmit},~\cite{roh2006design}. The complexity of the VQ algorithm increases (exact complexity analysis is shown in Sec.~\ref{sec::ComplexityAnalysis}) with increasing Tx antennas that makes the design algorithm impractical in massive MIMO regime. 

In this section, we focus on designing beamformer codebooks for the system model described in \ref{sec::BF} i.e., $M_r = 1$ and ${\rm rank}-1$ transmission. The UPA structure of the Tx antenna naturally allows us to represent the channel ${\bf H} \in {\C}^{1 \times M_t}$ as a matrix channel $\tilde{\bf H} \in {\C}^{M_v \times M_h}$ whose $(i,j)$-th element corresponds to the channel between the antenna element at the $i$-th row and $j$-th column of the UPA and the receive antenna. We first describe the design of unquantized beamformer for a given ${\bf H}$ and then provide a design method to construct the product codebooks for beamformer.

\subsection{Unquantized Beamformer Design}\label{sec::UnquantBeamformer}
The relation between the UPA matrix channel $\tilde{\bf H} \in {\C}^{M_v \times M_h}$ and ${\bf H} \in {\C}^{1 \times M_t}$ is ${\bf H}^T = {\rm vec}(\tilde{\bf H}^T)$. The SVD of $\tilde{\bf H}$ is $\tilde{\bf H} = {\tilde{\bf U}}{\tilde{\bf \Sigma}}{\tilde{\bf V}}^H$, where ${\tilde{\bf U}} \in {\cal U}_{M_v}$, ${\tilde{\bf V}} \in {\cal U}_{M_h}$, ${\tilde{\bf \Sigma}}$ is the $M_v \times M_h$ rectangular diagonal matrix with $i$-th largest singular value $\tilde{\sigma}_i$ at the entry ${(i, i)}$. Then we have
\begin{align}
    {\bf H}^T = {\rm vec}({\tilde{\bf H}}^T) = {\rm vec}({\tilde{\bf V}}^*{\tilde{\bf \Sigma}}{\tilde{\bf U}}^T) = {\rm vec} \left( \sum^{{\rm rank}(\tilde{\bf H})}_{i=1} {\tilde{\sigma}}_i{\tilde{\bf v}}_i^*{\tilde{\bf u}}^T_i \right) = \sum^{{\rm rank}(\tilde{\bf H})}_{i=1} {\tilde{\sigma}}_i{\tilde{\bf u}}_i \otimes {\tilde{\bf v}}^*_i. \label{eq::h::decomp}
\end{align}

Thus, we can represent ${\bf H}$ as the linear combination of ${\tilde{\bf u}}^T_i \otimes {\tilde{\bf v}}^H_i$ scaled with ${\tilde{\sigma}}_i$ as ${\bf H} = \displaystyle \sum^{{\rm rank}(\tilde{\bf H})}_{i=1} {\tilde{\sigma}}_i{\tilde{\bf u}}^T_i \otimes {\tilde{\bf v}}^H_i$. In order to facilitate product beamformer codebook construction, we approximate the channel ${\bf H}$ with its dominant direction, i.e., ${\tilde{\bf u}}^T_1 \otimes {\tilde{\bf v}}^H_1$, which is called the ${\rm rank}-1$ approximation. The approximated channel $\bar{\bf H}$ is given as ${\bf H} \approx \bar{\bf H} = {\tilde{\sigma}_1}{\tilde{\bf u}}^T_1 \otimes {\tilde{\bf v}}^H_1$. Let ${\bf f} \in \orthvect$ be a beamformer for $\bar{\bf H}$, then the KP form of $\bar{\bf H}$ naturally leads us to the idea of using ${\bf f}$ of the form ${\bf f} = {\bf f}_v \otimes {\bf f}_h$ where ${\bf f}_v \in {\cal U}(M_v, 1)$, ${\bf f}_h \in {\cal U}(M_h, 1)$. The beamforming gain ${\Gamma}(\bar{\bf H}, {\bf f})$ can now be simplified as ${\Gamma}(\bar{\bf H}, {\bf f}) = \norm{\bar{\bf H}{\bf f}}^2_2 =  \norm{{\tilde{\sigma}_1}({\tilde{\bf u}}^T_1 \otimes {\tilde{\bf v}}^H_1)({\bf f}_v \otimes {\bf f}_h)}^2_2 = {\tilde{\sigma}}^2_1 \norm{{\tilde{\bf u}}^T_1{\bf f}_v}_2^2\ \norm{{\tilde{\bf v}}^H_1{\bf f}_h}_2^2 =  {\tilde{\sigma}}^2_1\ |{\tilde{\bf u}}^T_1{\bf f}_v|^2\ |{\tilde{\bf v}}^H_1{\bf f}_h|^2$. The optimal beamformer $\hat{\bf f}$ for $\bar{\bf H}$ that maximizes ${\Gamma}(\bar{\bf H}, {\bf f})$ can be simplified as $\hat{\bf f} = \underset{{\bf f} \in {\orthvect}} \argmax\ { \Gamma(\bar{\bf H}, {\bf f})}$
\begin{align}
 =\underset{{\substack{{\bf f}_v \in {\cal U}(M_v,1) \\ {\bf f}_h \in {\cal U}(M_h,1)}}} \argmax\ |{\tilde{\bf u}}^T_1{\bf f}_v|^2\ |{\tilde{\bf v}}^H_1{\bf f}_h|^2 = \underset{{\bf f}_v \in {\cal U}(M_v,1)} \argmax\ |{\tilde{\bf u}}^T_1{\bf f}_v|^2 \otimes \underset{{\bf f}_h \in {\cal U}(M_h,1)} \argmax\ |{\tilde{\bf v}}^H_1{\bf f}_h|^2 = \hat{\bf f}_v \otimes \hat{\bf f}_h, \label{eq::prod::codeword}
\end{align}
where $\hat{\bf f}_v = \underset{{\bf f}_v \in {\cal U}(M_v,1)} \argmax\ |{\tilde{\bf u}}^T_1{\bf f}_v|^2, \ \hat{\bf f}_h =  \underset{{\bf f}_h \in {\cal U}(M_h,1)} \argmax\ |{\tilde{\bf v}}^H_1{\bf f}_h|^2$ and the maximum beamforming gain is ${ \Gamma(\bar{\bf H}, \hat{\bf f})} = \tilde{\sigma}_1^2$.  Clearly, a solution for the optimal beamformer $\hat{\bf f} = \hat{\bf f}_v \otimes \hat{\bf f}_h$ in \eqref{eq::prod::codeword} is given by the dominant singular vectors of the approximated channel $\tilde{\bf H}$, i.e., $\hat{\bf f}_v = {\tilde{\bf u}}^*_1$, $\hat{\bf f}_h = {\tilde{\bf v}}_1$ and thus $\hat{\bf f} = {\tilde{\bf u}}^*_1 \otimes {\tilde{\bf v}}_1$.

\subsection{Quantized Beamformer Design}\label{sec::QuantBeamformer}
We define the normalized beamforming gain ${\Gamma}_n(\bar{\bf H}, {\bf f})$ and the loss in ${\Gamma}_n(\bar{\bf H}, {\bf f})$, i.e., $L(\bar{\bf H}, {\bf f})$  obtained with an arbitrary KP beamformer ${\bf f} = {\bf f}_v \otimes {\bf f}_h$ as 
\begin{align}
    {\Gamma}_n(\bar{\bf H}, {\bf f}) &:= \frac{{\Gamma}(\bar{\bf H}, {\bf f})}{{\Gamma}(\bar{\bf H}, \hat{\bf f})} = \frac{{\Gamma}(\bar{\bf H}, {\bf f})}{{\tilde{\sigma}^2_1}} \stackrel{(a)}{=} |{\tilde{\bf u}}^T_1{\bf f}_v|^2\ |{\tilde{\bf v}}^H_1{\bf f}_h|^2, 
    L(\bar{\bf H}, {\bf f}) := 1 - {\Gamma}_n(\bar{\bf H}, {\bf f}), \notag 
\end{align}
where $\hat{\bf f}$ is the optimal unquantized KP beamformer for a given $\bar{\bf H}$, $(a)$ comes from \eqref{eq::prod::codeword}. The KP structure of the beamformer ${\bf f}$ motivates to employ separate codebooks ${\cal F}_v \subseteq {\cal U}(M_v, 1)$, ${\cal F}_h \subseteq {\cal U}(M_v, 1)$ for horizontal and vertical dimensions which enables to design product codebooks by clustering in lower dimensional spaces. The product codebook for the KP beamformer ${\bf f} = {\bf f}_v \otimes {\bf f}_h$ formed by the codebooks ${\cal F}_v$, ${\cal F}_h$ is represented as ${\cal F} = {\cal F}_v \times {\cal F}_h$. The loss in normalized beamforming gain with ${\bf f}$ can be bounded as $ L(\bar{\bf H}, {\bf f}) = 1 - {\Gamma}_n(\bar{\bf H}, {\bf f}) = 1 - |(\tilde{\bf u}^T_1 \otimes \tilde{\bf v}^H_1)({\bf f}_v\otimes {\bf f}_h)|^2 \leq$
\begin{align}
& 2\ \left( 1 - |(\tilde{\bf u}^T_1 \otimes \tilde{\bf v}^H_1)({\bf f}_v\otimes {\bf f}_h)| \right) \leq  2\ \underset{\theta, \phi}\min \left( \norm{(e^{j\theta}\tilde{\bf u}^*_1 \otimes e^{j\phi}\tilde{\bf v}_1) - ({\bf f}_v \otimes {\bf f}_h)} \right) \notag\\
&\leq 2\ \underset{\theta, \phi}\min \left(\norm{e^{j\theta}\tilde{\bf u}^*_1}_2\norm{e^{j\phi}\tilde{\bf v}_1 - {\bf f}_h}_2 + \norm{e^{j\theta}\tilde{\bf u}^*_1 - {\bf f}_v}_2\norm{e^{j\phi}{\bf f}_h}_2 \right)  \notag\\
&= 2\ \underset{\theta, \phi}\min \left(\norm{e^{j\phi}\tilde{\bf v}_1 - {\bf f}_h}_2 + \norm{e^{j\theta}\tilde{\bf u}^*_1 - {\bf f}_v}_2 \right) = 2\ \big[ (1 - |\tilde{\bf v}^H_1{\bf f}_h|)^{1/2} + (1 - |\tilde{\bf u}^T_1{\bf f}_v|)^{1/2} \big]  \notag\\
&\leq 2\ \big[(1 - |\tilde{\bf v}^H_1{\bf f}_h|^2) + (1 - |\tilde{\bf u}^T_1{\bf f}_v|^2)\big]  :=  L_{\rm ub}(\bar{\bf H}, {\bf f}). \notag 
\end{align}

In $L_{\rm ub}(\bar{\bf H}, {\bf f})$ defined above, for any angles $\alpha, \beta \in [0, 2\pi)$, we have $(1 - |\tilde{\bf v}^H_1{\bf f}_h|^2) + (1 - |\tilde{\bf u}^T_1{\bf f}_v|^2) = (1 - |\tilde{\bf v}^H_1{\bf f}_h{e^{j\alpha}}|^2) + (1 - |\tilde{\bf u}^T_1{\bf f}_v{e^{j\beta}}|^2)$. The rotational invariance of $L_{\rm ub}(\bar{\bf H}, {\bf f})$ from the above equation implies that ${\bf f}_v$, ${\bf f}_h$ are points on a GM i.e., ${\bf f}_v \in {\cal G}(M_v, 1)$, ${\bf f}_h \in {\cal G}(M_h, 1)$ and thus the respective codebooks ${\cal F}_v \subseteq {\cal G}(M_v, 1)$, ${\cal F}_h \subseteq {\cal G}(M_h, 1)$. From the definition of chordal distance $d_c(\cdot)$, the upper bound of $L(\bar{\bf H}, {\bf f})$ can also be written as
\begin{align}
    L_{\rm ub}(\bar{\bf H}, {\bf f}) &=  (1 - |\tilde{\bf v}^H_1{\bf f}_h|^2) + (1 - |\tilde{\bf u}^T_1{\bf f}_v|^2) =  d^2_c ({\tilde{\bf u}}^*_1, {\bf f}_v) + d^2_c ({\tilde{\bf v}}_1, {\bf f}_h). \notag
\end{align}

\begin{remark}
The upper bound of the loss in normalized beamforming gain i.e., $L_{\rm ub}(\bar{\bf H}, {\bf f})$ obtained by beamforming with ${\bf f} = {\bf f}_v \otimes {\bf f}_h$ instead of the optimal unquantized beamformer $\hat{\bf f} = {\tilde{\bf u}}^*_1 \otimes {\tilde{\bf v}}_1$ for a given ${\bf H}$ is equivalent to the squared distance between the points $({\tilde{\bf u}}^*_1, \tilde{\bf v}_1)$ and $({\bf f}_v, {\bf f}_h)$ on the CPM ${\cal G}^{\times}\left((M_v,M_h), (1,1)\right)$ i.e., $L_{\rm ub}(\bar{\bf H}, {\bf f}) = d^2_c ({\tilde{\bf u}}^*_1, {\bf f}_v) + d^2_c ({\tilde{\bf v}}_1, {\bf f}_h) =  d^2_c \left(({\tilde{\bf u}}^*_1, {\tilde{\bf v}}_1), ({\bf f}_v, {\bf f}_h) \right)$.
\end{remark}

\subsection{Product Codebook Design Criterion}
To measure the average distortion introduced by the quantization with the codebook ${\cal F} = {\cal F}_v \times {\cal F}_h$, we use the upper bound of the average loss in normalized beamforming gain $L_{\rm ub}(\bar{\bf H}, {\bf f})$ and define $L_{\rm ub}{(\cal F)}$ as $L_{\rm ub}{(\cal F)} := \E_{\bar{\bf H}}\ \big[L_{\rm ub}(\bar{\bf H}, {\bf f}) \big] = \E_{\tilde{\bf u}_1, \tilde{\bf v}_1}\ \big[L_{\rm ub}(\bar{\bf H}, {\bf f}) \big]$.

\begin{ndef}[Grassmann product codebook for beamforming] \label{def::grassmann::BF::productcodebook}
Under ${\rm rank}-1$ approximation of the channel, ${\bf H} \approx \bar{\bf H} = {\tilde{\sigma}}_1 \tilde{\bf u}^T_1 \otimes \tilde{\bf v}^H_1$, the Grassmann product codebook $\hat{\cal F} = \hat{\cal F}_v \times \hat{\cal F}_h$ for beamforming is the one that minimizes $L_{\rm ub}({\cal F})$ for a given feedback bit allocation $[B_v, B_h]$ where $|\hat{\cal F}_v| = 2^{B_v}$, $|\hat{\cal F}_h| = 2^{B_h}$. 
\end{ndef}
We will now state the method to construct the Grassmann product codebook $\hat{\cal F}$ as follows.

\begin{lemma}\label{lemm::BF::prod::codebook::design}
The Grassmann product codebook $\hat{\cal F} = \hat{\cal F}_v \times \hat{\cal F}_h$ as defined in Def.~\ref{def::grassmann::BF::productcodebook} can be constructed using the set of centroids ${\cal F}^K_v, {\cal F}^K_h$ obtained from the independent $K$-means clustering of the optimal KP beamformers $\tilde{\bf u}^*_1$, $\tilde{\bf v}_1$ on ${\cal G}(M_v,1)$, ${\cal G}(M_h,1)$ with $K = 2^{B_v}, 2^{B_h}$, respectively. 
\end{lemma}
\begin{IEEEproof}
See Appendix~\ref{app::BF::prod::codebook::design}.
\end{IEEEproof}

\subsection{Codebook construction}\label{sec::BF::DatasetConstruction}
From Lem.~\ref{lemm::BF::prod::codebook::design}, it is possible to perform $K$-means clustering independently on ${\cal G}(M_v,1)$, ${\cal G}(M_h,1)$ and construct the product codebook with reduced complexity. We assume a stationary distribution of the channel for a given coverage area of a Tx. In order to construct the Grassmann product codebook for beamforming as defined in Def.~\ref{def::grassmann::BF::productcodebook}, we construct ${\cal H} = \{{\bf H}\}$, a set of channel realizations sampled for different user locations. The available channel dataset ${\cal H}$ is split into training and testing datasets, ${\cal H}_{\rm train}$ and ${\cal H}_{\rm test}$ for generating beamformer codebooks and evaluating their performance respectively. We assume that the size of the training set is large enough so that the sampling distribution closely approximates the original distribution. The training procedure yields the optimal product codebook whose performance is evaluated by measuring the average normalized beamforming gain for the channel realizations in the test set ${\cal H}_{\rm test}$. The training and testing procedure of the proposed product codebook design for a given set of channel realizations is summarized in the following remark.

\begin{remark}
For a given ${\cal H}_{\rm train}$ and ${\cal H}_{\rm test}$, the Grassmann product codebook for beamforming $\hat{\cal F} = \hat{\cal F}_v \times \hat{\cal F}_h$ is obtained by the procedure {\sc BFTrain}$({\cal H}_{\rm train}$,$[B_v, B_h])$ and the performance of the codebook $\hat{\cal F}$ is evaluated by the procedure {\sc BFTest}$({\cal H}_{\rm test}$,$[\hat{\cal F}_v, \hat{\cal F}_h])$ as outlined in Alg.~\ref{alg::BF::TrainingTesting::Algo}, where $B_v$, $B_h$ are the number of bits used to encode $\tilde{\bf u}^*_1$, $\tilde{\bf v}_1$ respectively.
\end{remark}

\section{Product Codebook Design for Precoding}
In this section, we present a product codebook design method for ${\rm rank}-\texttt{r}$ $(M_h > \texttt{r}, M_v > \texttt{r})$ transmission in a MIMO system with $M_r > 1$ as described in Sec.~\ref{sec::Precoding}. Similar to the beamformer codebook design, we explore the UPA structure of the Tx antenna and tensor representation of the channel to find reduced complexity precoder codebooks. We introduce this scheme as follows.

\subsection{HOOI-based Unquantized Precoder Design}
\subsubsection{Tucker decomposition of the channel}
The uniform planar structure of the Tx antenna permits a natural representation of the matrix channel ${\bf H}$ as tensor $\nscrH$ where $\nscrH \in \C^{M_r\times M_h\times M_v}$ (as demonstrated in Fig.~\ref{fig::Tensor::Channel}) and ${\nscrH}_{ijk}$ represents the channel between the antenna element at $k$-th row and $j$-th column of the UPA at the Tx and the $i$-th antenna at the Rx. Although one can rearrange ${\bf H}$ in tensors of arbitrary dimensions, in the rest of this paper, we will be focusing on the tensors of dimensions $M_r\times M_h\times M_v$. From the tensor representation of channel ${\bf H}$ as ${\nscrH}$, we have that ${\bf H}$ is equivalent to the mode-$1$ unfolding of ${\nscrH}$ i.e., ${\bf H} = {\bf H}_{(1)}$ and TD of ${\nscrH}$ is expressed as
\begin{align}
   {\nscrH} &= {\nscrG} \times_1 {\bf B} \times_2 {\bf A}^{(1)} \times_3 {\bf A}^{(2)}, {\bf H} = {\bf H}_{(1)} = {\bf B}{\bf G}_{(1)}({\bf A}^{(2)} \otimes {\bf A}^{(1)})^{T}  = {\bf B}{\bf G}_{(1)}{\bf A}^{H}, \notag 
\end{align}
where ${\nscrG} \in {\C}^{M_r \times M_h \times M_v}$ is the core tensor, ${\bf B} \in {\cal U}_{M_r}$, ${\bf A}^{(1)} \in {\cal U}_{M_h}$, ${\bf A}^{(2)} \in {\cal U}_{M_v}$,${\bf A} = \left({\bf A}^{(2)} \otimes {\bf A}^{(1)}\right)^{*}$. The best ${\rm rank}-({M_r,\texttt{r},\texttt{r}})$ approximation of ${\nscrH}$ i.e., ${\bar{\nscrH}}$ obtained as described in Sec.~\ref{sec::Tensors} is
\begin{align}
   {\nscrH} \approx \bar{\nscrH} &= \bar{\nscrG} \times_1 \bar{\bf B} \times_2 \bar{\bf A}^{(1)} \times_3 \bar{\bf A}^{(2)}, \bar{\bf H} = \bar{\bf H}_{(1)} = \bar{\bf B}\bar{\bf G}_{(1)}(\bar{\bf A}^{(2)} \otimes \bar{\bf A}^{(1)})^{T}  = \bar{\bf B}\bar{\bf G}_{(1)}\bar{\bf A}^{H}, \label{eq::TenDecompLR::Unfolding}
\end{align}
where $\bar{\nscrG} \in {\C}^{M_r \times \texttt{r} \times \texttt{r}}$ is the core tensor, $\bar{\bf B} \in {\cal U}_{M_r}$, $\bar{\bf A}^{(1)} \in {\cal U}({M_h, \texttt{r}})$, $\bar{\bf A}^{(2)} \in {\cal U}({M_v, \texttt{r}})$, $\bar{\bf A} = \left(\bar{\bf A}^{(2)} \otimes \bar{\bf A}^{(1)}\right)^{*}$. Here, $\bar{\bf H}$ is the mode-$1$ unfolding of the $\bar{\nscrH}$ and $\bar{\bf A}^{(1)}$, $\bar{\bf A}^{(2)}$ are the principal components of $\bar{\nscrH}$ in the horizontal, vertical dimensions, respectively.

From the SVD of channel ${\bf H}$, the eigenvalue $\sigma^2_i$ represents the power of the channel along the corresponding eigen-direction ${\bf v}_i$. We recall that in SVD-based precoding, an optimal precoder for ${\rm rank}-\texttt{r}$ transmission is formed by dominant $\texttt{r}$ columns of ${\bf V}$ i.e., the columns of ${\bf V}$ corresponding to the dominant $\texttt{r}$ singular values. The basic principle of the proposed HOOI-based precoder design technique is also to identify the dominant $\texttt{r}$ columns of $\bar{\bf A} = \left(\bar{\bf A}^{(2)} \otimes \bar{\bf A}^{(1)}\right)^{*}$ in \eqref{eq::TenDecompLR::Unfolding} that maximize the mutual information when the ${\rm rank}-\texttt{r}$ matrix formed by the $\texttt{r}$ columns is used as precoder for transmission. However, identifying the dominant $\texttt{r}$ columns of $\bar{\bf A}$ out of ${\texttt{r}}^2$ columns is not immediately clear, since unlike the singular matrix ${\bf \Sigma}$, $\bar{\bf G}_{(1)}$ is not a diagonal matrix. Let ${\cal C} \subset \{1,\cdots,{\texttt{r}}^2\}$ with $|{\cal C}|  = \texttt{r}$ be a set of column indices and ${\cal C}_o$ be the set of column indices of dominant $\texttt{r}$ columns of $\bar{\bf A}$ and $\bar{\bf A}_{\cal C} = \bar{\bf A}(:,{\cal C})$. The construction of ${\cal C}_o$ and the proposed unquantized precoder for a given ${\bf H}$ are outlined as follows.

\begin{prop}\label{Prop::UnquantPrecoder}
For a given ${\bf H}$, the proposed unquantized precoder for ${\rm rank}-\texttt{r}$ transmission is formed by the dominant $\texttt{r}$ columns of $\bar{\bf A}$ i.e., $\bar{\bf A}_{{\cal C}_o}$, where ${\cal C}_o$ is the set of column indices of dominant $\texttt{r}$ columns of $\bar{\bf A}$ that maximizes the mutual information $R({\bf H},\bar{\bf A}_{{\cal C}_o})$. 
\end{prop}
The mutual information obtained with the precoder $\bar{\bf A}_{\cal C}$ for a given ${\bf H}$ is $R({\bf H},\bar{\bf A}_{\cal C}) = \log{\rm det}\left({\bf I} + \rho_t\bar{\bf A}^H_{\cal C}{\bf H}^H{\bf H}\bar{\bf A}_{\cal C}\right)$. Then, ${\cal C}_o$ is obtained from the following optimization problem:
\begin{align}
    {\cal C}_o =  \underset{{\cal C} \subset \{1,\cdots,{\texttt{r}}^2\}, |{\cal C}|  = \texttt{r}} \argmax R({\bf H},\bar{\bf A}_{{\cal C}}) &=  \underset{{\cal C} \subset \{1,\cdots,{\texttt{r}}^2\}, |{\cal C}|  = \texttt{r}} \argmax \log{\rm det}\left({\bf I} + \rho_t\bar{\bf A}^H_{\cal C}{\bf H}^H{\bf H}\bar{\bf A}_{\cal C}\right) \notag\\ 
    &=  \underset{{\cal C} \subset \{1,\cdots,{\texttt{r}}^2\}, |{\cal C}|  = \texttt{r}} \argmax  \log{\rm det}\left({\bf I} + \rho_t({\bf H}\bar{\bf A}_{\cal C})^H({\bf H}\bar{\bf A}_{\cal C})\right). \label{eq::Unquant::PropPrecoder}
\end{align}
The above optimization is equivalent to choosing the appropriate $\texttt{r}$ columns out of ${\texttt{r}}^2$ columns of ${\bf H}{\bar{\bf A}}$ and the exact solution ${\cal C}_o$ is obtained by maximizing $R({\bf H},\bar{\bf A}_{{\cal C}})$ over all the possible $\texttt{r}$ element sets for ${\cal C}$. Interestingly, $R({\bf H},\bar{\bf A}_{{\cal C}})$ is a monotone submodular function~\cite{konar2017greed} and hence \eqref{eq::Unquant::PropPrecoder} is a monotone submodular maximization problem with cardinality constraints (see Sec.~\ref{sec::Submodularity}). Since this problem is NP hard~\cite{konar2017greed}, we provide a greedy algorithm in Alg.~\ref{alg::DomColumns} for the design of ${\cal C}_o$.

\begin{lemma}\label{cor::RateMaximization}
The mutual information obtained with the proposed unquantized precoder $\bar{\bf A}_{{\cal C}_o}$ is $R({\bf H},\bar{\bf A}_{{\cal C}_o})  = R(\bar{\bf H},\bar{\bf A}_{{\cal C}_o}) = \log{\rm det}\left({\bf I} + \rho_t \bar{\bf G}^H_{(1),{\cal C}_o}\bar{\bf G}_{(1),{\cal C}_o} \right)$.
\end{lemma}
\begin{IEEEproof}
Consider the equivalent channel ${\bf H}_{eq}$ associated with the precoder $\bar{\bf A}_{{\cal C}}$ and ${\bf H}$. Then, we have $
   {\bf H}_{eq}^H{\bf H}_{eq} = \bar{\bf A}^H_{{\cal C}}{\bf H}_{(1)}^H{\bf H}_{(1)}\bar{\bf A}_{{\cal C}} = \bar{\bf A}^H_{{\cal C}}\bar{\bf H}_{(1)}^H\bar{\bf H}_{(1)}\bar{\bf A}_{{\cal C}} = \bar{\bf A}^H_{{\cal C}}\bar{\bf A}\bar{\bf G}_{(1)}^H \bar{\bf G}_{(1)} \bar{\bf A}^H \bar{\bf A}_{{\cal C}} = \bar{\bf G}^H_{(1),{\cal C}}\bar{\bf G}_{(1),{\cal C}} 
$. From Alg.~\ref{alg::DomColumns}, the proposed unquantized precoder can be expressed as $\bar{\bf A}_{{\cal C}_o} = \textsc{DomCol}(\bar{\bf A}, {\bf H}, \texttt{r})$ and thus the mutual information  is $
    R({\bf H},\bar{\bf A}_{{\cal C}_o}) 
    = \underset{{{{\cal C} \subseteq \{1,\cdots,{\texttt{r}}^2\},   |{\cal C}|  = \texttt{r}}}} \max R({\bf H},\bar{\bf A}_{{\cal C}}) = \log{\rm det}\left({\bf I} + \rho_t\bar{\bf A}^H_{{\cal C}_o}{\bf H}^H{\bf H}\bar{\bf A}_{{\cal C}_o} \right)  
    = \log{\rm det}\left({\bf I} + \rho_t \bar{\bf G}^H_{(1),{\cal C}_o}\bar{\bf G}_{(1),{\cal C}_o} \right) 
     = R(\bar{\bf H},\bar{\bf A}_{{\cal C}_o}). 
$
\end{IEEEproof}

In optimal precoding, the Tx requires the knowledge of $\bar{\bf V}$. Whereas, in HOOI-based precoding, the Tx requires the knowledge of $\bar{\bf A}_{{\cal C}_o}$ which is formed using $\bar{\bf A}^{(1)}$, $\bar{\bf A}^{(2)}$ and ${\cal C}_o$ as described in Lem.~\ref{cor::RateMaximization}. As the channel realization ${\bf H}$ changes, $\bar{\bf A}^{(1)}$, $\bar{\bf A}^{(2)}$ change and ${\cal C}_o$ that forms the proposed precoder $\bar{\bf A}_{{\cal C}_o}$ also changes. Hence, for this scheme, $(\bar{\bf A}^{(1)}, \bar{\bf A}^{(2)}, {\cal C}_o)$ is the CSIT required for the construction of the precoder. However, due to the limited capacity of the feedback channel, this information needs to be quantized.

\subsection{Quantized Precoder Design}\label{sec::codebook::design}

In this section, we propose the design of quantized precoder and a loss in mutual information due to quantization for a given ${\bf H}$ that enable the design of product precoder codebooks, which are cartesian product of two lower dimensional codebooks. The KP structure of $\bar{\bf A} = (\bar{\bf A}^{(2)} \otimes \bar{\bf A}^{(1)})^*$ in the precoder $\bar{\bf A}_{{\cal C}_o}$ motivates to construct a ${\rm rank}-\texttt{r}$ precoder of the form $\left(Q(\bar{\bf A})\right)_{{\cal C}_Q}$, where $Q(\bar{\bf A}) = \left(Q(\bar{\bf A}^{(2)}) \otimes Q(\bar{\bf A}^{(1)})\right)^*$, and $Q(\bar{\bf A}^{(1)}) \in {\cal U}(M_h, \texttt{r})$, $Q(\bar{\bf A}^{(2)}) \in {\cal U}(M_v, \texttt{r})$ are the quantized versions of $\bar{\bf A}^{(1)}$, $\bar{\bf A}^{(2)}$, respectively, ${\cal C}_Q$ is a set of $\texttt{r}$ column indices of $Q(\bar{\bf A})$. On the similar lines of design of unquantized precoder in Prop.~\ref{Prop::UnquantPrecoder}, ${\cal C}_Q$ is designed to maximize the mutual information with the precoder $\left(Q(\bar{\bf A})\right)_{{\cal C}_Q}$. We formally describe the construction of the optimal quantized precoder in the following proposition.
 
\begin{prop}\label{Prop::QuantPrecoder}
Let $Q(\bar{\bf A}^{(1)}) \in {\cal U}(M_h,\texttt{r})$ and $Q(\bar{\bf A}^{(2)}) \in {\cal U}(M_v,\texttt{r})$ be the quantized representations of $\bar{\bf A}^{(1)}$ and $\bar{\bf A}^{(2)}$ respectively. Then, for a given ${\bf H}$, the proposed quantized precoder for ${\rm rank}-\texttt{r}$ transmission is formed by the dominant $\texttt{r}$ columns of $Q(\bar{\bf A}) = \left(Q(\bar{\bf A}^{(2)}) \otimes Q(\bar{\bf A}^{(1)})\right)^*$ i.e., $\left(Q(\bar{\bf A})\right)_{{\cal C}_Q}$, where ${\cal C}_Q$ is the set of column indices of dominant $\texttt{r}$ columns of $Q(\bar{\bf A})$ which maximizes $R\left({\bf H}, \left(Q(\bar{\bf A})\right)_{{\cal C}_Q}\right)$.
\end{prop}
The mutual information obtained with the precoder $\left(Q(\bar{\bf A})\right)_{\cal C}$ for a given ${\bf H}$ is $R\left({\bf H},\left(Q(\bar{\bf A})\right)_{\cal C}\right) = \log{\rm det}\left({\bf I} +\rho_t \left(Q(\bar{\bf A})\right)^H_{\cal C}{\bf H}^H{\bf H}\left(Q(\bar{\bf A})\right)_{\cal C}\right)$. From Prop.~\ref{Prop::QuantPrecoder}, ${\cal C}_Q$ is obtained as
\begin{align}
    {\cal C}_{Q} &=  \underset{{\substack{{\cal C} \subseteq \{1,\cdots,{\texttt{r}}^2\} \\ |{\cal C}|  = \texttt{r}}}} \argmax R\left({\bf H},\left(Q(\bar{\bf A})\right)_{\cal C}\right) =  \underset{{\substack{{\cal C} \subseteq \{1,\cdots,{\texttt{r}}^2\} \\ |{\cal C}|  = \texttt{r}}}} \argmax  \log{\rm det}\left({\bf I} + \rho_t\left(Q(\bar{\bf A})\right)^H_{\cal C}{\bf H}^H{\bf H}\left(Q(\bar{\bf A})\right)_{\cal C}\right). \label{eq::Quant::PropPrecoder}
\end{align}
The above optimization corresponds to maximizing a monotone submodular function with cardinality constraints similar to \eqref{eq::Unquant::PropPrecoder}. The exact solution for ${\cal C}_Q$ is obtained by maximizing $R\left({\bf H},\left(Q(\bar{\bf A})\right)_{{\cal C}}\right)$ over all the possible $\texttt{r}$ element sets for ${\cal C}$ which is NP-hard to determine. Thus, the proposed optimal quantized precoder can be expressed as $\left(Q(\bar{\bf A})\right)_{{\cal C}_Q} = \textsc{DomCol}\left(Q(\bar{\bf A}), {\bf H}, \texttt{r} \right)$ (refer to Alg.~\ref{alg::DomColumns}). With the quantized principal components $Q(\bar{\bf A}^{(1)})$, $Q(\bar{\bf A}^{(2)})$ and ${\cal C}_Q$, the Tx is able to construct $\left(Q(\bar{\bf A})\right)_{{\cal C}_Q}$ for precoding. 

To measure the average loss in mutual information due to the limited capacity of the feedback channel, we first define a loss in mutual information associated with an arbitrary precoder ${\bf F} \in {\cal U}(M_t,\texttt{r})$ for a given ${\bf H}$ as $L(\bar{\bf H},{\bf F}) : = R(\bar{\bf H},\bar{\bf A}_{{\cal C}_o}) - R(\bar{\bf H},{\bf F})$ where
\begin{align}
    R(\bar{\bf H},{\bf F}) &=  \log{\rm det}\left({\bf I} + \rho_t{\bf F}^H\bar{\bf H}^H\bar{\bf H}{\bf F}\right) = \log{\rm det}\left({\bf I} + \rho_t{\bf F}^H{\bar{\bf A}}{\bar{\bf G}^H_{(1)}}{\bar{\bf G}_{(1)}}{\bar{\bf A}^H}{\bf F}\right) \notag \\ 
    &\stackrel{>}{\sim} \log{\rm det}\left({\bf I} + \rho_t{\bf F}^H{\bar{\bf A}}_{{\cal C}_o}{\bar{\bf G}^H_{(1),{{\cal C}_o}}}{\bar{\bf G}_{(1),{{\cal C}_o}}}{\bar{\bf A}^H}_{{\cal C}_o}{\bf F}\right) : = R_{\rm lb}(\bar{\bf H},{\bf F}). \label{eq::Rate::LB}
\end{align}
For concise notation let ${\bar{\bf G}^H_{(1),{{\cal C}_o}}}{\bar{\bf G}_{(1),{{\cal C}_o}}} = \bar{\bf \Lambda}_{{\cal C}_o}$, then
\begin{align}
 R_{\rm lb}(\bar{\bf H},{\bf F}) = \log{\rm det}\left({\bf I} + \rho_t\bar{\bf \Lambda}_{{\cal C}_o}\right) + \log{\rm det}\big[{\bf I} - ({\bf I} + \rho_t\bar{\bf \Lambda}_{{\cal C}_o})^{-1} \rho_t\bar{\bf \Lambda}_{{\cal C}_o} \left({\bf I} - {\bar{\bf A}^H}_{{\cal C}_o}{\bf F}{\bf F}^H{\bar{\bf A}}_{{\cal C}_o}\right)\big], \label{eq::Rate::LBmodified}
\end{align}
since $({\bf I} +\rho_t\bar{\bf \Lambda}_{{\cal C}_o}{\bar{\bf A}^H}_{{\cal C}_o}{\bf F}{\bf F}^H{\bar{\bf A}}_{{\cal C}_o}) = \big[({\bf I} +\rho_t{\bar{\bf \Lambda}_{{\cal C}_o}}) - \rho_t\bar{\bf \Lambda}_{{\cal C}_o}({\bf I} - {\bar{\bf A}^H}_{{\cal C}_o}{\bf F}{\bf F}^H{\bar{\bf A}}_{{\cal C}_o})\big]$. $L(\bar{\bf H},{\bf F})$ can be bounded as
\begin{align}
    L(\bar{\bf H},{\bf F}) &: = R(\bar{\bf H},\bar{\bf A}_{{\cal C}_o}) - R(\bar{\bf H},{\bf F})
    \stackrel{(a)}{\leq} R(\bar{\bf H},\bar{\bf A}_{{\cal C}_o}) - R_{\rm lb}(\bar{\bf H},{\bf F}) \notag \\
    &\leq  \log{\rm det}\left[{\bf I} - ({\bf I} + \rho_t\bar{\bf \Lambda}_{{\cal C}_o})^{-1} \rho_t\bar{\bf \Lambda}_{{\cal C}_o} \left({\bf I} - {\bar{\bf A}^H}_{{\cal C}_o}{\bf F}{\bf F}^H{\bar{\bf A}}_{{\cal C}_o}\right)\right] :=   L_{\rm ub}(\bar{\bf H},{\bf F}), \label{eq::LossRate::UB}
\end{align}
where $(a)$ is obtained from \eqref{eq::Rate::LBmodified}. Because of the difficulty in directly working with the upper bound of loss, we approximate $L_{\rm ub}(\bar{\bf H},{\bf F})$ under high-resolution (number of feedback bits $B$ is reasonably large) and high-SNR $(\rho_t \rightarrow \infty)$ approximations. When the number of feedback bits $B$ (high-resolution) are large, we have that ${\bar{\bf A}^H}_{{\cal C}_o}{\bf F}{\bf F}^H{\bar{\bf A}}_{{\cal C}_o}$ is close to ${\bf I}$ and when $\rho_t$ is large, $({\bf I} + \rho_t\bar{\bf \Lambda}_{{\cal C}_o})^{-1} \rho_t\bar{\bf \Lambda}_{{\cal C}_o} \approx {\bf I}$. Therefore $L_{\rm ub}(\bar{\bf H},{\bf F})$ can be further approximated as 
\begin{align}
     L_{\rm ub}(\bar{\bf H},{\bf F}) &\stackrel{\text{large B}}{\approx} {\rm tr}\left(({\bf I} + \rho_t\bar{\bf \Lambda}_{{\cal C}_o})^{-1} \rho_t\bar{\bf \Lambda}_{{\cal C}_o} \left({\bf I} - {\bar{\bf A}^H}_{{\cal C}_o}{\bf F}{\bf F}^H{\bar{\bf A}}_{{\cal C}_o}\right)\right) \stackrel{\text{high } \rho_t}{\approx} {\rm tr}({\bf I} - {\bar{\bf A}^H}_{{\cal C}_o}{\bf F}{\bf F}^H{\bar{\bf A}}_{{\cal C}_o}).  \label{eq::HighRes::HighSNR::Loss::approx}
\end{align}
In the next section, we use the above defined loss for designing the low-complexity product precoder codebooks. 

\subsection{Product Codebook Design Criterion}
Let ${\cal F}_{h} \subseteq {\cal U}(M_h,\texttt{r})$, ${\cal F}_{v} \subseteq {\cal U}(M_v,\texttt{r})$ be the codebooks to quantize $\bar{\bf A}^{(1)}$, $\bar{\bf A}^{(2)}$, respectively. Then the codebook ${\cal F}$ corresponding to $\bar{\bf A}$ is constructed using ${\cal F}_{h}$ and ${\cal F}_{v}$ as below. 
\begin{align}
    {\cal F} = \{\left({\bf F}_{v} \otimes {\bf F}_{h}\right)^*\}\ \forall\ {\bf F}_{h} \in {\cal F}_{h}, {\bf F}_{v} \in {\cal F}_{v}.
\end{align}
Therefore ${\cal F} \subseteq {\cal U}(M_t, \texttt{r}^2)$ and precisely, ${\cal F}$ is a finite collection of orthonormal matrices from the tensor product space ${\cal U}(M_h, \texttt{r})$ and ${\cal U}(M_v, \texttt{r})$ i.e., ${\cal F} \subseteq {\cal U}(M_v, \texttt{r}) \otimes {\cal U}(M_h, \texttt{r})$. The mapping of $\bar{\bf A}^{(1)}$, $\bar{\bf A}^{(2)}$ to the appropriate codewords from ${\cal F}_{h}$, ${\cal F}_{v}$ can be represented as $Q: {\cal U}(M, \texttt{r}) \mapsto {\cal F}$, where $(M, {\cal F}) = (M_h, {\cal F}_{h})$, $(M, {\cal F}) = (M_v, {\cal F}_{v})$ for $\bar{\bf A}^{(1)}$, $\bar{\bf A}^{(2)}$, respectively and thus the quantized $\bar{\bf A}$ is obtained as $Q(\bar{\bf A}) = \left(Q(\bar{\bf A}^{(2)}) \otimes Q(\bar{\bf A}^{(1)})\right)^*$. As we proceed, we design the optimal codebooks $\hat{\cal F}, \hat{\cal F}_{h}$, $\hat{\cal F}_{v}$ and the quantizer mapping $Q(\cdot)$ such that average distortion due to quantization is minimized.

From \eqref{eq::HighRes::HighSNR::Loss::approx}, the average of the defined loss in mutual information with precoder $\left(Q(\bar{\bf A})\right)_{{\cal C}_Q}$ is
 \begin{align}
 \E_{\bf H}\left[L\left(\bar{\bf H},\left(Q(\bar{\bf A})\right)_{{\cal C}_Q} \right)\right] &= \E_{\bf H}\left[{\rm tr}\left({\bf I} - \left(Q(\bar{\bf A})\right)^H_{{\cal C}_Q}{\bar{\bf A}}_{{\cal C}_o}{\bar{\bf A}}^H_{{\cal C}_o} \left(Q(\bar{\bf A})\right)_{{\cal C}_Q}\right) \right], \label{eq::AvgLoss::General}
 \end{align}
and the optimal codebook $\hat{\cal F}$ that minimizes the above average loss is 
\begin{align}
 \hat{\cal F} &= \underset{{\cal F} \subseteq {\cal U}(M_v, \texttt{r}) \otimes {\cal U}(M_h, \texttt{r})} \argmin\ \underset{Q(\cdot)} \min\  \E_{\bf H}\left[L\left(\bar{\bf H},\left(Q(\bar{\bf A})\right)_{{\cal C}_Q} \right)\right] \notag \\
 &= \underset{{\cal F} \subseteq {\cal U}(M_v, \texttt{r}) \otimes {\cal U}(M_h, \texttt{r})} \argmin\ \underset{Q(\cdot)} \max\  \E_{\bf H}\left[{\rm tr}\left(\left( Q(\bar{\bf A})\right)^H_{{\cal C}_Q}{\bar{\bf A}}_{{\cal C}_o}{\bar{\bf A}}^H_{{\cal C}_o} \left(Q(\bar{\bf A})\right)_{{\cal C}_Q}\right) \right] \notag \\
&= \underset{{\cal F} \subseteq {\cal U}(M_v, \texttt{r}) \otimes {\cal U}(M_h, \texttt{r})} {\argmax}\ \underset{Q(\cdot)} \max\  \E_{{\bf H}} \left[ \norm{\bar{\bf A}^H_{{\cal C}_o}{\left(Q(\bar{\bf A})\right)_{{\cal C}_Q}}}^2_F \right]. \notag 
\end{align}

For every ${\bf H}$, the set of indices of $\texttt{r}$ dominant columns of the unquantized and quantized precoder i.e., ${\cal C}_o$ and ${\cal C}_{Q}$ change. To enable the product codebook structure and de-tangle the maximization objective, instead of maximizing $\E_{{\bf H}}\left[\norm{\bar{\bf A}^H_{{\cal C}_o}{\left(Q(\bar{\bf A})\right)_{{\cal C}_Q}}}^2_F\right]$ for designing the codebooks, $\E_{{\bf H}}\left[ \norm{\bar{\bf A}^H Q(\bar{\bf A})}^2_F \right]$ is maximized. Thus the codebook design criterion is modified as
\begin{align}
   \hat{\cal F}  &=  \underset{{{{\cal F} \subseteq  {\cal U}(M_v, \texttt{r}) \otimes {\cal U}(M_h, \texttt{r})}}} \argmax\ \underset{Q(\cdot)} \max\ \E_{{\bf H}}\left[\norm{\left(\bar{\bf A}^{(2)} \otimes \bar{\bf A}^{(1)}\right)^H\left(Q(\bar{\bf A}^{(2)}) \otimes Q(\bar{\bf A}^{(1)})\right)}^2_F\right]. \label{eq::DesignCriterion::Unitary}
\end{align} 

\subsection{Connection with Product Grassmann Manifold}
In the above objective, for any ${\rm rank}-\texttt{r}$ unitary matrices ${\bf Q}_1, {\bf Q}_2 \in {\cal U}_{\texttt{r}}$ we have
\begin{align}
    \norm{\left(\bar{\bf A}^{(2)} \otimes \bar{\bf A}^{(1)}\right)^H\left(Q(\bar{\bf A}^{(2)}) \otimes Q(\bar{\bf A}^{(1)})\right)}^2_F &= \norm{\left(\bar{\bf A}^{(2)} \otimes \bar{\bf A}^{(1)} \right)^H\left(Q(\bar{\bf A}^{(2)}){\bf Q}_2 \otimes Q(\bar{\bf A}^{(1)}){\bf Q}_1 \right)}^2_F. \notag 
\end{align}
It follows that $\norm{\left(\bar{\bf A}^{(2)} \otimes \bar{\bf A}^{(1)}\right)^H\left(Q(\bar{\bf A}^{(2)}) \otimes Q(\bar{\bf A}^{(1)})\right)}^2_F$ should be maximized not just over orthonormal matrices in ${\cal U}(M_v, \texttt{r}) \otimes {\cal U}(M_h, \texttt{r})$ but over equivalence classes of such matrices i.e., over all the matrices such that $Q(\bar{\bf A}^{(1)}){\bf Q}_1 \sim Q(\bar{\bf A}^{(1)})$ and $Q(\bar{\bf A}^{(2)}){\bf Q}_2 \sim Q(\bar{\bf A}^{(2)})$. This means that \eqref{eq::DesignCriterion::Unitary} should be maximized over GMs. Therefore the codebooks ${\cal F}$, ${\cal F}_{h}$ and ${\cal F}_{v}$ can be interpreted as collection of orthonormal basis of subspaces in the GMs i.e., ${\cal F}_{h} \subseteq {\cal G}(M_h, \texttt{r})$, and ${\cal F}_{v} \subseteq {\cal G}(M_v, \texttt{r})$ and thus ${\cal F} \subseteq {\cal G}(M_v, \texttt{r}) \otimes {\cal G}(M_h, \texttt{r})$. Similar to a CPM ${\cal G}^{\times}\left((M_v,M_h), (\texttt{r},\texttt{r})\right)$, ${\cal G}(M_v, \texttt{r}) \otimes {\cal G}(M_h, \texttt{r})$ represents another type of product manifold known as TPM. The $m$-{\em {fold TPM}} is the subset ${\cal G}^{\otimes}({\bf n}, {\bf k}) := \{{\bf F}_1 \otimes \cdots \otimes {\bf F}_m | {\bf F}_i \in {\cal G}(n_i, k_i), i = 1,\cdots,m\} \subset {\cal G}(N, K)$,  where $({\bf n}, {\bf k}) := \left((n_1, k_1), (n_2,k_2), \cdots, (n_m,k_m)\right)$,  $N = n_1n_2\cdots n_m, K = k_1k_2\cdots k_m$. The following lemma draws a relation between the two product manifolds, TPM and CPM. 

\begin{lemma}\label{lem::Diffeomorphism}
The m-fold TPM ${\cal G}^{\otimes}({\bf n}, {\bf k})$ is diffeomorphic to the m-fold CPM ${\cal G}^{\times}({\bf n}, {\bf k})$ i.e., the map $\varphi: {\cal G}^{\times}({\bf n}, {\bf k}) \mapsto {\cal G}^{\otimes}({\bf n}, {\bf k})$ is a diffeomorphism\footnote{The existence of diffeomorphism between the two manifolds ${\cal G}^{\otimes}({\bf n}, {\bf k})$ and ${\cal G}^{\otimes}({\bf n}, {\bf k})$ implies that the map $\varphi$ is bijective, $\varphi, \varphi^{-1}$ are smooth, continuous, and differentiable as well. See~\cite{curtef2012riemannian} for a more rigorous discussion.}. 
\end{lemma}

Hence, there exists a one-to-one mapping from any point ${\bf F}_1 \otimes \cdots \otimes {\bf F}_m \in {\cal G}^{\otimes}({\bf n}, {\bf k})$ to $({\bf F}_1, \cdots, {\bf F}_m) \in {\cal G}^{\times}({\bf n}, {\bf k})$ and vice-versa. Now we provide an approximation for $d_c(\cdot)$ on ${\cal G}^{\otimes}({\bf n}, {\bf k})$ which will be used in constructing the proposed product precoder codebooks.

\begin{assumption}\label{rem::ChordalDist::Approx}
If ${\bf F}_1 \otimes \cdots \otimes {\bf F}_m$, ${\bf F}'_1 \otimes \cdots \otimes {\bf F}'_m $ are any two points on ${\cal G}^{\otimes}({\bf n}, {\bf k})$, then their preimages on ${\cal G}^{\times}({\bf n}, {\bf k})$ are $[\bf F] = ({\bf F}_1,\cdots, {\bf F}_m)$, $[\bf F'] = ({\bf F}'_1,\cdots, {\bf F}'_m)$, respectively. We approximate the distance between the points on the TPM with the distance between their preimages on the CPM as $d^2_c\left({\bf F}_1 \otimes \cdots \otimes {\bf F}_m, {\bf F}'_1 \otimes \cdots \otimes {\bf F}'_m\right) \approx d^2_c\left([\bf F], [\bf F']\right) \approx \sum\limits_{i=1}^m d^2_c({\bf F}_i, {\bf F}_i')$.
\end{assumption}

The codebook design criterion in \eqref{eq::DesignCriterion::Unitary} can be interpreted using $d_c (\cdot)$ defined on a GM and can be modified as
 $\hat{\cal F} =  \underset{{\cal F} \subseteq  {\cal G}^{\otimes}\left((M_v,M_h), (\texttt{r},\texttt{r})\right)} \argmin\ \underset{Q(\cdot)}  \min \ \E_{{\bf H}}\big[ d^2_c{\left(\bar{\bf A}, Q(\bar{\bf A})\right)} \big]$.  
Therefore, the objective for designing the optimal codebook $\hat{\cal F}$ is equivalent to minimizing the average chordal distance between the two points $\left(\bar{\bf A}^{(2)} \otimes \bar{\bf A}^{(1)}\right)$ and $\left(Q(\bar{\bf A}^{(2)}) \otimes Q(\bar{\bf A}^{(1)})\right)$ on ${\cal G}^{\otimes}\left((M_v,M_h), (\texttt{r},\texttt{r})\right)$. From the diffeomorphism between the TPM ${\cal G}^{\otimes}\left((M_v,M_h), (\texttt{r},\texttt{r})\right)$ and the CPM ${\cal G}^{\times}\left((M_v,M_h), (\texttt{r},\texttt{r})\right)$, the above optimization objective for $\hat{\cal F}$ has the following equivalent statement.
\begin{align}
 \hat{\cal F}  &=   \underset{{\cal F} \subseteq  {\cal G}^{\times}\left((M_v,M_h), (\texttt{r},\texttt{r})\right)} \argmin\ \underset{Q(\cdot)}  \min \ \E_{{\bf H}}\left[d^2_c\left(\left(\bar{\bf A}^{(2)}, \bar{\bf A}^{(1)}\right), \left(Q(\bar{\bf A}^{(2)}), Q(\bar{\bf A}^{(1)})\right)\right)\right] \label{eq::DesignCriterion::GrassmannEq}
\end{align} 

Also, the minimization objective in the above design criterion can be regarded as a measure of average loss in mutual information with a codebook ${\cal F}$, where $Q(\bar{\bf A}^{(1)}) \in {\cal F}_{h}$, $Q(\bar{\bf A}^{(2)}) \in {\cal F}_{v}$ and thus $L_{\rm ub}({\cal F}) = \E_{{\bf H}}\left[d^2_c\left(\left(\bar{\bf A}^{(2)}, \bar{\bf A}^{(1)}\right), \left(Q(\bar{\bf A}^{(2)}), Q(\bar{\bf A}^{(1)})\right)\right)\right]$.

\begin{ndef}[Grassmann product codebook for precoding] \label{def::grassmann::PC::productcodebook}
Under the ${\rm rank}-(M_r, \texttt{r}, \texttt{r})$ approximation of the channel, ${\bf H} \approx \bar{\bf H}_{(1)} = \bar{\bf B}\bar{\bf G}_{(1)}(\bar{\bf A}^{(2)} \otimes \bar{\bf A}^{(1)})^{T}$, the Grassmann product codebook $\hat{\cal F} = \hat{\cal F}_{v} \times \hat{\cal F}_{h}$ for precoding is the one that minimizes $L_{\rm ub}({\cal F})$ for a given feedback bit allocation $[B_v, B_h]$ where $|\hat{\cal F}_{h}| = 2^{B_h}$, $|\hat{\cal F}_{v}| = 2^{B_v}$. 
\end{ndef}

We now state the method to construct $\hat{\cal F}$ as follows.

\begin{lemma}\label{lemm::PC::prod::codebook::design}
The Grassmann product codebook $\hat{\cal F} = \hat{\cal F}_{v} \times \hat{\cal F}_{h}$ as defined in Def.~\ref{def::grassmann::PC::productcodebook} can be constructed using the set of centroids ${\cal F}^K_{h}$, ${\cal F}^K_{v}$ obtained from the independent $K$-means clustering of the principal components $\bar{\bf A}^{(1)}$, $\bar{\bf A}^{(2)}$ on ${\cal G}(M_h,\texttt{r})$, ${\cal G}(M_v,\texttt{r})$ with $K = 2^{B_h}, 2^{B_v}$, respectively. 
\end{lemma}
\begin{IEEEproof}
See Appendix~\ref{app::pc::prod::codebook::design}.
\end{IEEEproof}
\begin{remark}
The design criterion for optimal product codebook in \eqref{eq::Kmeans::simplification} is equivalent to finding the set of optimal $K$ centroids using the $K$-means clustering algorithm on the CPM ${\cal G}^{\times}\left((M_v,M_h), (\texttt{r},\texttt{r})\right)$ with the chordal distance metric induced on a CPM. The relation between the chordal distance between two points on a CPM and its factor manifolds as given in \eqref{eq::ChordalDistance::DPGM} simplifies the objective to two separate objectives of finding the optimal centroids using $K$-means clustering algorithm on the factor manifolds of the CPM ${\cal G}^{\times}\left((M_v,M_h), (\texttt{r},\texttt{r})\right)$. 
\end{remark}
The step-wise construction of the proposed unquantized and quantized precoders is summarized in the following remark. 
\subsection{Codebook Construction}
From Lem.~\ref{lemm::PC::prod::codebook::design}, it is possible to perform $K$-means clustering independently on ${\cal G}(M_v,\texttt{r})$, ${\cal G}(M_h,\texttt{r})$ and construct the product precoder codebook with reduced complexity. The construction of the training and testing channel datasets ${\cal H}_{\rm train}$ and ${\cal H}_{\rm test}$ for precoder codebook design is similar to the construction provided for beamforming product codebook design in Sec.~\ref{sec::BF::DatasetConstruction}. The training procedure yields the optimal precoder codebooks whose performance is evaluated by measuring the average mutual information $R_{\rm av}$ for the channel realizations in the test set ${\cal H}_{\rm test}$ obtained with the proposed quantized precoder construction. The training and testing procedure of the codebook design for a given set of channel realizations is given in the following remark.

\begin{remark}
For a given ${\cal H}_{\rm train}$ and ${\cal H}_{\rm test}$, the Grassmann product codebook for precoding $\hat{\cal F} = \hat{\cal F}_{v} \times \hat{\cal F}_{h}$ is obtained by the procedure {\sc PCTrain}$({\cal H}_{\rm train}$,$[B_v, B_h])$ and the performance of the codebook $\hat{\cal F}$ is evaluated by the procedure {\sc PCTest}$({\cal H}_{\rm test}$,$[\hat{\cal F}_{v}, \hat{\cal F}_{h}])$ as outlined in Alg.~\ref{alg::PC::TrainingTesting::Algo}, where $B_h$, $B_v$ are the number of bits used to encode $\bar{\bf A}^{(1)}$, $\bar{\bf A}^{(2)}$ respectively. 
\end{remark}

\begin{figure*}
\begin{minipage}[t]{0.48\textwidth}
\begin{algorithm}[H]
\centering
\caption{\small Grassmannian $K$-means Algorithm }\label{alg::Product::KmeansAlgo}
\scriptsize
\begin{algorithmic}[1]
\Procedure{Codebook}{${{\cal X}},[K, n, k]$}
\State Initialize random ${\cal F} = \{{\bf F}_1,\cdots,{\bf F}_K\}$ on ${\cal G}(n, k)$
\State  \textbf{Cluster Update}: ${\cal S}_i \leftarrow \{{\bf X}: d_c({\bf X},{\bf F}_i) \leq d_c({\bf X},{\bf F}_j), \forall {\bf X} \in {\cal X}, \ i \neq j \}\ \forall i \in \{1,\cdots,K\}$ 
\State  \textbf{Quantization}: $Q_{\cal F}({\bf X}$) $\leftarrow$ $\underset{{\bf F} \in {\cal F}}\argmin\ d_c^2({\bf X},{\bf F})\ \forall {\bf X} \in {\cal X}$ 
\While{$\text{!~stopping criteria}$}
\State \textbf{Centroid Update}: ${\bf F}_i \leftarrow \underset{{\bf F} \in {\cal G}(n, k)} \argmin \sum d_c^2({\bf X},{\bf F})\ \forall {\bf X} \in {\cal S}_i, \forall i \in \{1,\cdots, K\}$
\State \textbf{Cluster Update} \text{ and } \textbf{Quantization} \notag
\EndWhile

\Return ${\cal F}$
\EndProcedure
\end{algorithmic}
\end{algorithm} 
\vspace{-15pt}

\begin{algorithm}[H]
    \centering
    \caption{\small Training, testing of the Grassmann product codebook for beamforming}\label{alg::BF::TrainingTesting::Algo}
  \scriptsize
    \begin{algorithmic}[1]
\Procedure{BFTrain}{$\cal H_{\rm train}$, $[B_v,B_h]$}
\State Initialize training sets ${\cal X}_{\rm train}=\emptyset$ and ${\cal Y}_{\rm train}=\emptyset$ on ${\cal G}(M_h,1)$ and ${\cal G}(M_v,1)$ respectively
\For{${\bf H}\in {\cal  H}_{\rm train}$}
\State Construct $\tilde{\bf H}$ from $\tilde{\bf H}$
\State ${\tilde{\bf U}}{\tilde{\bf \Sigma}}{\tilde{\bf V}}^H \leftarrow {\rm svd}(\tilde{\bf H})$
\State ${\cal X}_{\rm train}\leftarrow {\cal X}_{\rm train}\cup {\bf v}_1 $, ${\cal Y}_{\rm train}\leftarrow {\cal Y}_{\rm train}\cup {\bf u}^*_1 $
\EndFor 
\State $\hat{\cal F}_h \leftarrow$ {\sc CodeBook}(${\cal X}_{\rm train}$, $[2^{B_h}, M_h, 1]$)
\State $\hat{\cal F}_v \leftarrow$ {\sc CodeBook}(${\cal Y}_{\rm train}$, $[2^{B_v}, M_v, 1]$)

\Return $[\hat{\cal F}_v$, $\hat{\cal F}_h]$
\EndProcedure
\Procedure{BFTest}{$\cal H_{\rm test}$, $[\hat{\cal F}_v$, $\hat{\cal F}_h]$}
\State Initialize $\Gamma_{\rm av}=0$
\For{${\bf H} \in {\cal  H}_{\rm test}$}
\State Generate $\tilde{\bf H}$ from ${\bf H}$ 
\State $\tilde{\bf U}\tilde{\bf \Sigma}\tilde{\bf V}^H \leftarrow {\rm svd}(\tilde{\bf H})$
\State ${\bf f}_h \leftarrow \underset{{{\bf f} \in \hat{\cal F}_h}} \argmin\ d_c^2({\bf v}_1, {\bf f})$, 
\State ${\bf f}_v \leftarrow \underset{{{\bf f} \in \hat{\cal F}_v}} \argmin\ d_c^2({\bf u}^*_1, {\bf f})$
\State $\Gamma_{\rm av} \leftarrow  \Gamma_{\rm av}+\frac{1}{\#{\cal H}_{\rm test}} \frac{\Gamma({\bf H}, {\bf f}_v \otimes {\bf f}_h)}{\Gamma({\bf H}, {\bf u}^*_1 \otimes {\bf v}_1)}$
\EndFor

\Return ${\Gamma}_{\rm av}$
\EndProcedure
\end{algorithmic}
\end{algorithm}
\end{minipage}
\hfill
\begin{minipage}[t]{0.48\textwidth}
\begin{algorithm}[H]
\centering
\caption{\small Greedy algorithm to find the $\texttt{r}$ dominant columns that forms the precoder in \eqref{eq::Unquant::PropPrecoder} and \eqref{eq::Quant::PropPrecoder} for a given ${\bf H}$}\label{alg::DomColumns}
\scriptsize
\begin{algorithmic}[1]
\Procedure{DomCol}{${\bf X}, {\bf H}, \texttt{r}$}
\State Initialize ${\cal C}^1_o = \emptyset$, $i = 1$
\While{$i \leq \texttt{r}$}
\State $c_i = \underset{c_i \notin {\cal C}^{i-1}_o} \argmax \log{\rm det}\left({\bf I} +\rho_t ({\bf H}{\bf X}_{{\cal C}^{i-1}_o})({\bf H}{\bf X}_{{\cal C}^{i-1}_o})^H \right)$
\State ${\cal C}^i_o = {\cal C}^{i-1}_o \cup \{c_i\} $
\EndWhile
\State ${\cal C}_o \leftarrow {\cal C}^{\texttt{r}}_o$

\Return ${\bf X}_{{\cal C}_o}$
\EndProcedure
\end{algorithmic}
\end{algorithm} 

\vspace{-19pt}

\begin{algorithm}[H]
  \scriptsize
    \centering
    \caption{  \small Training, testing of the Grassmann product codebook for precoding}\label{alg::PC::TrainingTesting::Algo}
\begin{algorithmic}[1]
\Procedure{PCTrain}{$\cal H_{\rm train}$, $[B_v,B_h]$}
\State Initialize training sets ${\cal A}_{i, \rm train} = \emptyset$ and ${\cal A}_{2, \rm train} = \emptyset$ on ${\cal G}(M_h,\texttt{r})$ and ${\cal G}(M_v,\texttt{r})$ respectively
\For{${\bf H}\in {\cal  H}_{\rm train}$}
\State Construct ${\nscrH}$ from ${\bf H}$, $\bar{\nscrH}$ from ${\nscrH}$ 
\State $\bar{\bf B}\bar{\bf G}_{(1)}(\bar{\bf A}^{(2)} \otimes \bar{\bf A}^{(1)})^{T} \leftarrow \bar{\bf H}_{(1)}$
\State ${\cal A}_{i, \rm train}  \leftarrow {\cal A}_{i, \rm train}  \cup \bar{\bf A}^{(i)}$, $(i = 1,2)$ 
\EndFor 
\State $\hat{\cal F}_{j} \leftarrow$ {\sc CodeBook}$({\cal A}_{i, \rm train} $, $[2^{B_{j}}, M_{j}, r])$ $((i,j) = (1,h),(2,v))$

\Return $[\hat{\cal F}_{v}$, $\hat{\cal F}_{h}]$
\EndProcedure
\Procedure{PCTest}{$\cal H_{\rm test}$, $[\hat{\cal F}_{v}$, $\hat{\cal F}_{h}]$}
\State Initialize $R_{\rm av}=0$
\For{${\bf H} \in {\cal  H}_{\rm test}$}
\State Construct ${\nscrH}$ from ${\bf H}$ and $\bar{\nscrH}$ from ${\nscrH}$ 
\State $\bar{\bf B}\bar{\bf G}_{(1)}(\bar{\bf A}^{(2)} \otimes \bar{\bf A}^{(1)})^{T} \leftarrow \bar{\bf H}_{(1)}$
\State $Q(\bar{\bf A}^{(i)}) \leftarrow \underset{{{\bf F} \in \hat{\cal F}_{j}}} \argmin\ d_c^2(\bar{\bf A}^{(i)}, {\bf F}), \forall (i,j)$
\State $Q(\bar{\bf A})  \leftarrow \left( Q(\bar{\bf A}^{(2)}) \otimes Q(\bar{\bf A}^{(1)}) \right)^*$
\State $\left(Q(\bar{\bf A})\right)_{{\cal C}_Q} \leftarrow \textsc{DomCol}\left(Q(\bar{\bf A}), {\bf H}, \texttt{r} \right)$
\State $R_{\rm av} \leftarrow  R_{\rm av} + \frac{1}{\#{\cal H}_{\rm test}}R\left({\bf H}, \left(Q(\bar{\bf A})\right)_{{\cal C}_Q}\right)$
\EndFor
\Return $R_{\rm av}$
\EndProcedure
\end{algorithmic}
\end{algorithm}
\end{minipage}
\end{figure*}  

\section{Complexity Analysis}\label{sec::ComplexityAnalysis}
In this section, we compute and compare the complexity of the proposed product codebook design technique with the VQ based iterative codebook design method provided in \cite{roh2006transmit,roh2006design} using a detailed complexity analysis. Let the total number of points in the channel training dataset available for the codebook design be $N$, number of codewords in the codebook be $K$. Each iteration of the Grassmannian $K$-means clustering algorithm involves the following steps: the computation of pairwise distances between cluster centroids and data points and the computation of centroid of the data points that belong to each cluster and updating the codebook.

The distance $d_c({\bf X}, {\bf Y})$ between any two points ${\bf X}, {\bf Y} \in {\cal G}(M,\texttt{r})$ requires computation of SVD of ${\bf X}^H{\bf Y} \in {\C}^{\texttt{r} \times \texttt{r}}$ whose complexity is ${\cal O}({\texttt{r}}^3 + M{\texttt{r}}^2)$. Therefore the complexity of computing the distance between $K$ centroids and $N$ data points on ${\cal G}(M,\texttt{r})$ is ${\cal O}(KN{\texttt{r}}^3 + KNM{\texttt{r}}^2)$. For the calculation of centroid of a set of $p$ points belonging to a cluster according to Lem.~\ref{lem::Centroid::Computation}, it is required to compute SVD of an $M \times M$ matrix obtained by the sum of $p$ $M \times M$ matrices and hence the complexity is ${\cal O}(M^2\texttt{r}p+M^3)$. This gives the computational cost of calculation of $K$ centroids as ${\cal O}(M^2N\texttt{r} + KM^3)$. Thus the total computation cost for a single iteration of the Grassmannian $K$-means clustering algorithm on ${\cal G}(M, \texttt{r})$ is ${\cal O}(M^2N\texttt{r} + KM^3 + KN{\texttt{r}}^3 + KNM{\texttt{r}}^2)$.


For the iterative VQ design method in~\cite{roh2006design}, the set of optimal centroids of the ${\rm rank}-\texttt{r}$ right singular matrices $\bar{\bf V} \in {\C}^{M_t \times \texttt{r}}$ of the channel dataset ${\cal H}_{\rm train}$ forms the precoder codebook. This gives the complexity of single iteration of the VQ design method as ${\cal O}(M^2_tN\texttt{r} + KM^3_t + KN{\texttt{r}}^3 + KNM_t{\texttt{r}}^2)$. For the proposed product beamformer and precoder codebook design method, two codebooks with $K'$ codewords each, corresponding to horizontal and vertical dimensions have to constructed using Alg.~\ref{alg::BF::TrainingTesting::Algo} and~\ref{alg::PC::TrainingTesting::Algo}. The complexity of a single iteration of construction of $\hat{\cal F}_{h}$ from ${\cal A}_{1,{\rm train}}$ is ${\cal O}(M^2_h \texttt{r}N + K'M^3_h + K'NM_h^2{\texttt{r}}^2 + K'N{\texttt{r}}^3)$ and that of $\hat{\cal F}_{v}$ from ${\cal A}_{2,{\rm train}}$ is ${\cal O}(M^2_v \texttt{r}N + K'M^3_v + K'NM_v^2{\texttt{r}}^2 + K'N{\texttt{r}}^3)$.

\begin{remark}\label{Remark::ComplexityAnalysis}
Let $M_h = M_v = n$, then $M_t = n^2$ and the computational complexity of the VQ design method in \cite{roh2006design} is ${\cal O}(n^4N\texttt{r} + Kn^6 + KN{\texttt{r}}^3 + KNn^2{\texttt{r}}^2)$ whereas the proposed scheme has significantly lower complexity of ${\cal O}(2n^2\texttt{r}N + 2K'n^3 + 2K'N{\texttt{r}}^3 + 2K'Nn{\texttt{r}}^2)$ for ${\rm rank}-\texttt{r}$ transmission.
\end{remark}
In the massive MIMO regime, as $M_h$, $M_v$ increase, construction of codebooks with quartic complexity in~\cite{roh2006design} can become impractical whereas the proposed method with quadratic complexity is relatively computationally efficient. We will validate this fact with  numerical results presented next.

\section{Results and Discussions}\label{sec::Results}

\subsection{Dataset generation}\label{sec::dataset::gen}
For the performance evaluation of the Grassmann product codebooks, we consider an indoor communication scenario between the base station and the users operating at 2.5 GHz. The channel realizations are obtained from the DeepMIMO dataset~\cite{alkhateeb2019deepmimo}, which specifies the ray tracing channel parameters for different locations. The parameters for the generation of channel dataset are provided in Table.~\ref{dataset::table::Journal}.

\begin{table}
\small
\centering
\begin{tabular}{ |c|c| } 
 \hline
 Name of scenario & I1\_2p5 \\ 
  \hline
 Active BS & 3  \\ 
  \hline
 Active users & 1 to 702  \\ 
 \hline
Number of antennas (x, y, z) & ($M_v,M_h,M_r$) \\
 \hline
 System bandwidth & 0.02 GHz \\
 \hline
 Antennas spacing & 0.5 \\
 \hline
 Number of OFDM sub-carriers & 1 \\
 \hline
 OFDM sampling factor & 1 \\
 \hline
 OFDM limit & 1 \\
 \hline
\end{tabular}
\caption{Parameters of the DeepMIMO dataset~\cite{alkhateeb2019deepmimo}}\label{dataset::table::Journal}
\vspace{0.1in}
\end{table}
\begin{figure}[t]
\centering
\begin{subfigure}{0.32\textwidth}
  \centering
  \includegraphics[scale=0.37]{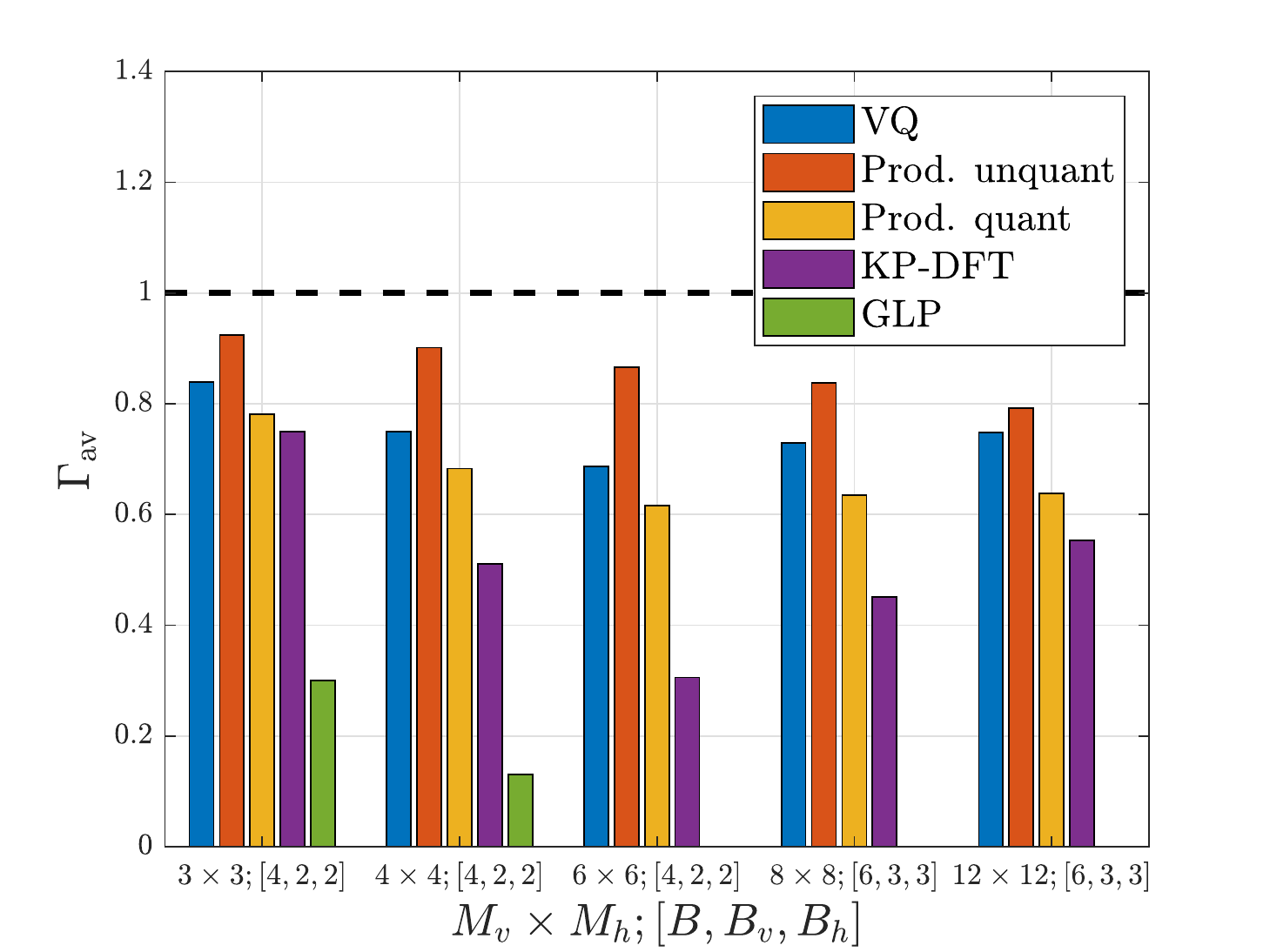}
\caption{}
\label{fig::Results::Rate::Mr1}
\end{subfigure}%
\hfill
\begin{subfigure}{0.32\textwidth}
  \centering
  \includegraphics[scale=0.37]{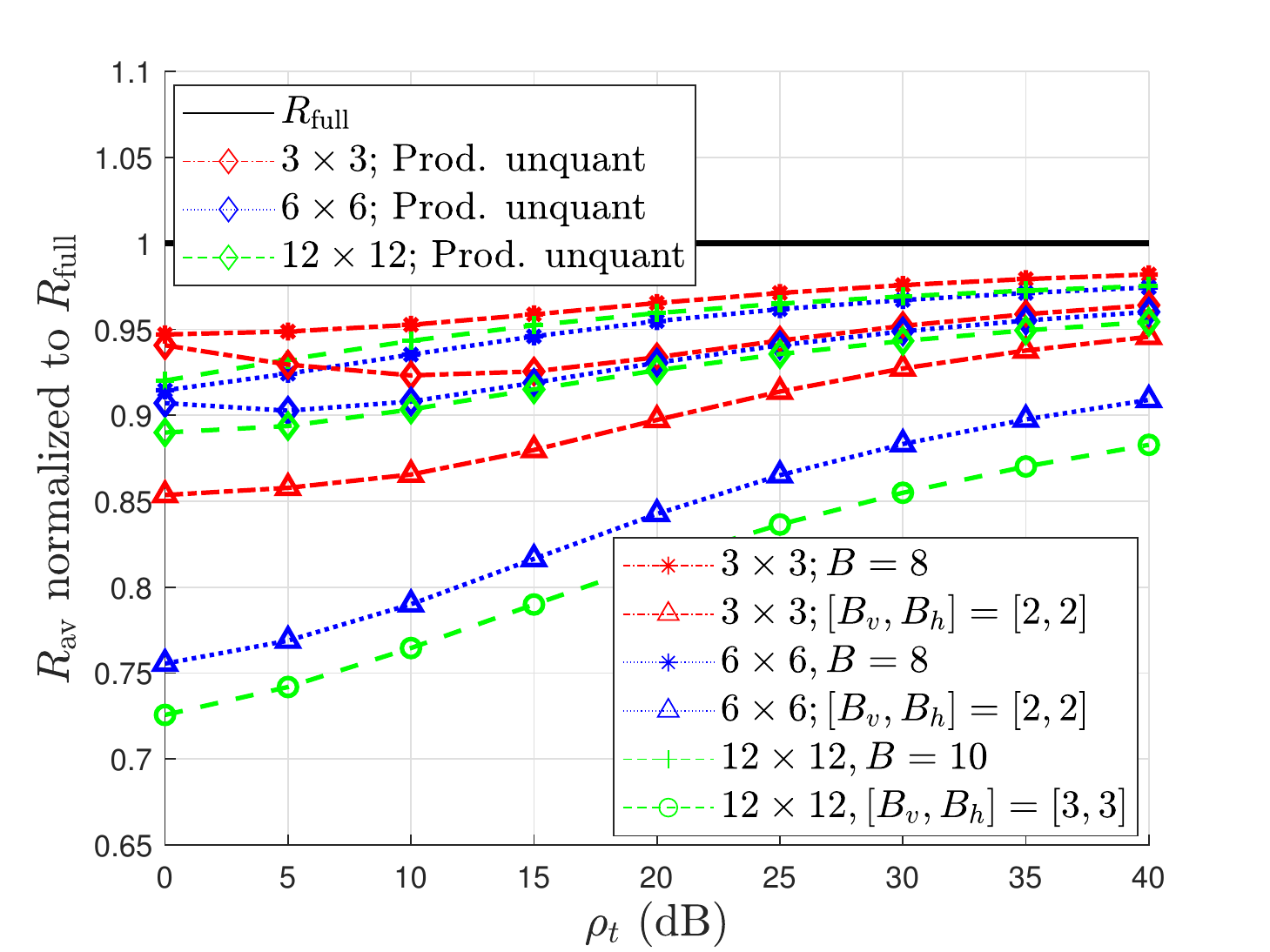}
 \caption{}
\label{fig::Results::Rate::Mr2sweep}
\end{subfigure}%
\hfill
\begin{subfigure}{0.32\textwidth}
  \centering
   \includegraphics[scale=0.37]{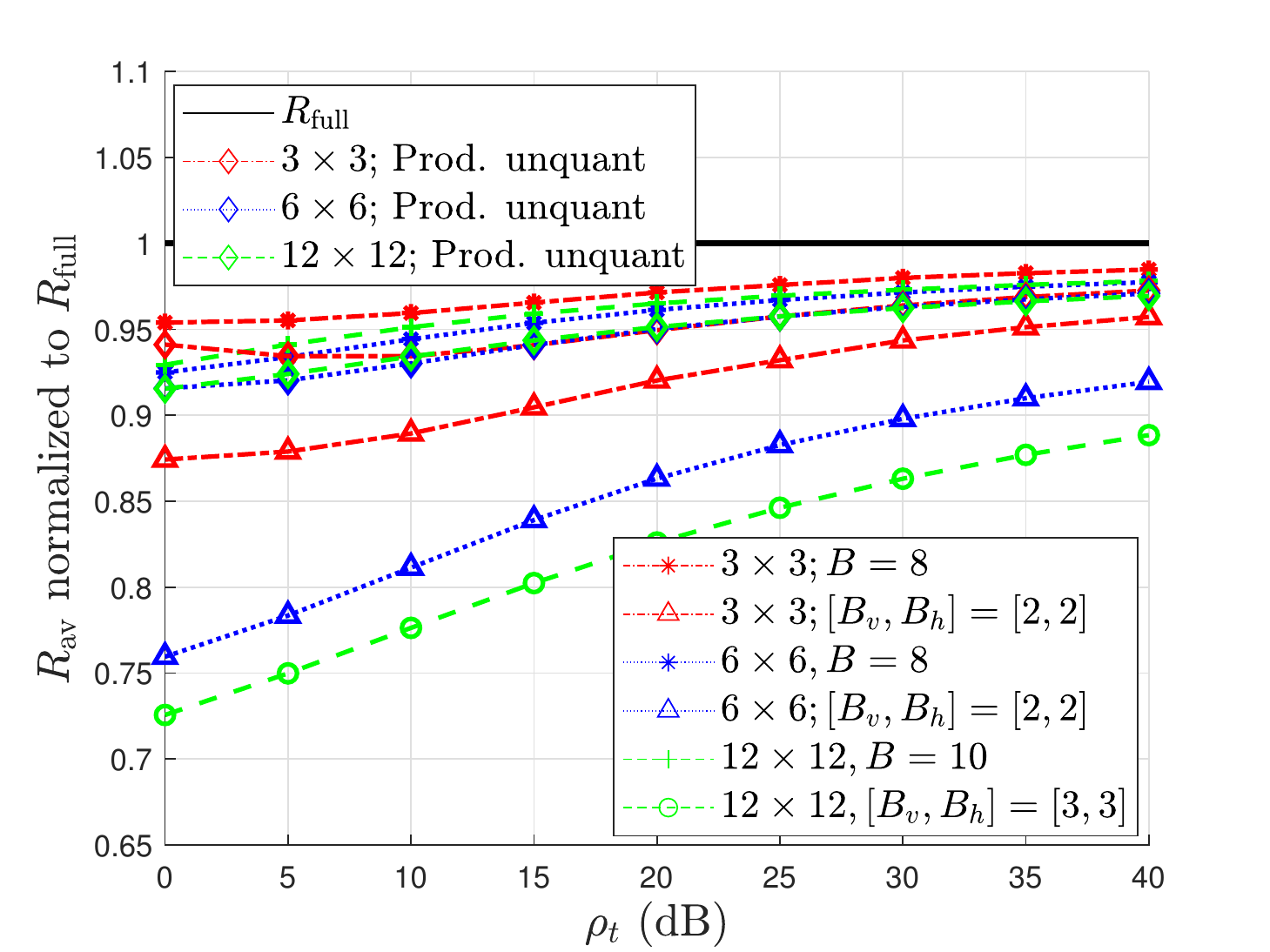}
  \caption{}
\label{fig::Results::Rate::Mr3sweep}
\end{subfigure}
\caption{Performance comparison of the proposed Grassmann product codebooks with VQ method~\cite{roh2006design} for various Tx antenna configurations $M_v \times M_h$ and feedback bit allocations $[B,B_v,B_h]$. (a) ${\Gamma}_{\rm av}$ for $M_r=1,  \texttt{r} = 1$, (b) $R_{\rm av}$ normalized to $R_{\rm full}$ for $M_r = 2, \texttt{r} = 2$ at varying $\rho_t$, and (c) $R_{\rm av}$ normalized to $R_{\rm full}$ for $M_r = 3, \texttt{r} = 2$ at varying $\rho_t$}
\label{fig::Results1}
\end{figure}
\subsection{Results}
We present numerical results to assess the performance of the designed product codebooks for beamforming and precoding in FD-MIMO systems in terms of ${\Gamma}_{\rm av}$ and $R_{\rm av}$, respectively. For a given Tx antenna configuration $M_v \times M_h$  and feedback bits allocation ({$[B,B_v,B_h]$}), the codebooks  are generated using  Lem.~\ref{lemm::BF::prod::codebook::design} and \ref{lemm::PC::prod::codebook::design}, respectively.  Here, $[B,B_v,B_h]$ denotes the feedback bit allocation for the limited feedback scheme where $B$ bits are used for the codebooks using the VQ method (referred to as `VQ')~\cite{roh2006design, roh2006transmit, bhogi2020learning} and $[B_v, B_h]$ is the feedback bit allocation for the Grassmann product codebooks (referred to as `Prod. quant'). To demonstrate the quantization loss, we also plot ${\Gamma}_{\rm av}$ and $R_{\rm av}$ for the unquantized beamformer and precoder (referred to as `Prod. unquant') as defined in Sec.~\ref{sec::UnquantBeamformer} and Prop.~\ref{Prop::UnquantPrecoder} respectively. 

\begin{figure}[t]
\centering
\begin{subfigure}{0.32\textwidth}
  \centering
  \includegraphics[scale=0.37]{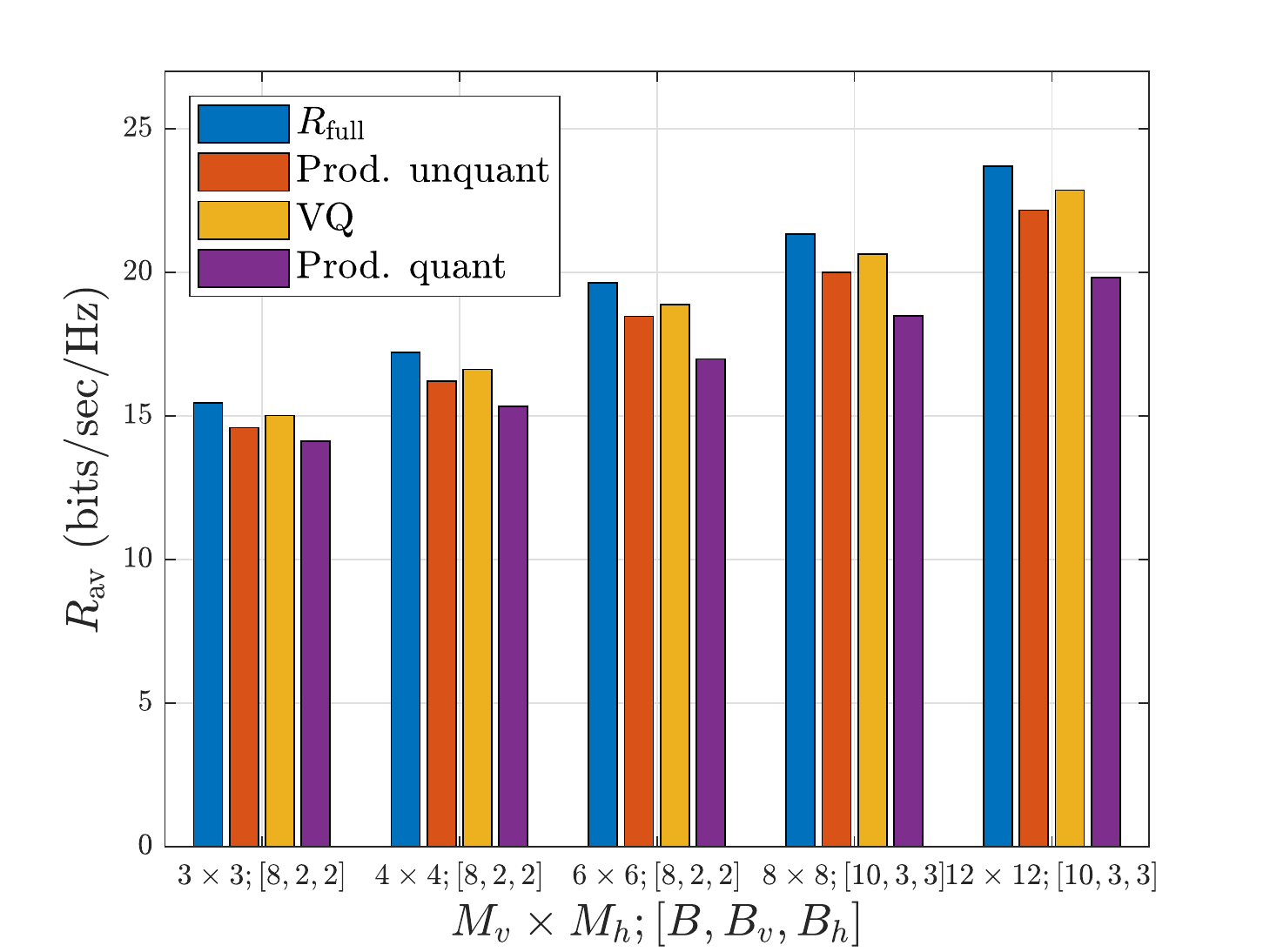}
  \caption{}
\label{fig::Results::Rate::Mr2}
\end{subfigure}%
\hfill
\begin{subfigure}{0.32\textwidth}
  \centering
  \includegraphics[scale=0.37]{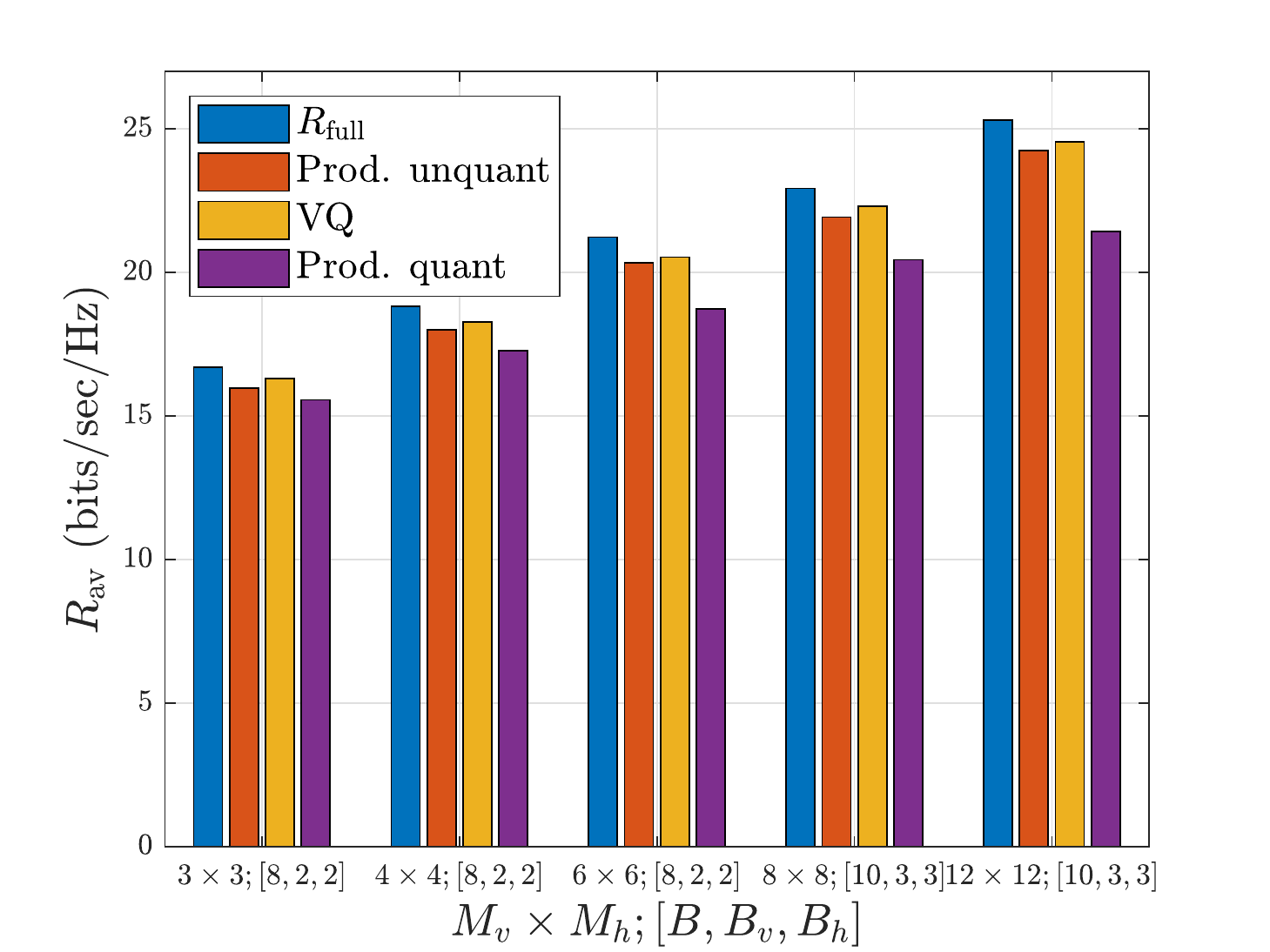}
 \caption{}
\label{fig::Results::Rate::Mr3}
\end{subfigure}%
\hfill
\begin{subfigure}{0.32\textwidth}
  \centering
   \includegraphics[scale=0.37]{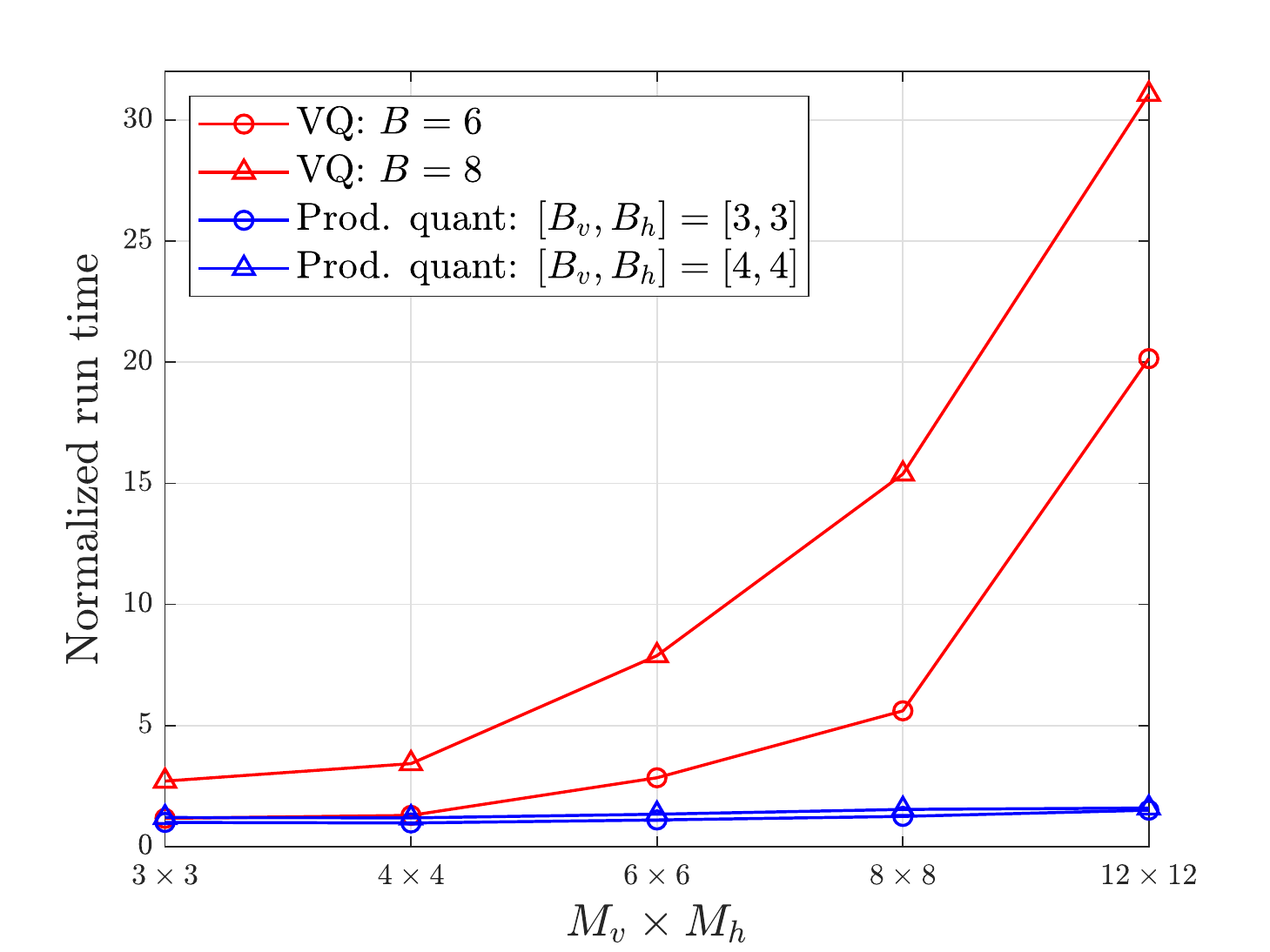}
 \caption{}
\label{fig::Results::RunTime}
\end{subfigure}
\caption{Performance comparison of the proposed Grassmann product codebooks with VQ method~\cite{roh2006design} for various Tx antenna configurations $M_v \times M_h$ and feedback bit allocations $[B,B_v,B_h]$. (a) $R_{\rm av}$ for $M_r = 2, \texttt{r} = 2, \rho_t = 25$ dB, (b) $R_{\rm av}$ for $M_r = 2, \texttt{r} = 3, \rho_t = 25$ dB, and (c) Normalized run-times for $M_r = 2, \texttt{r} = 2$ }
\label{fig::Results2}
\end{figure}

In Fig.~\ref{fig::Results::Rate::Mr1}, we compare ${\Gamma}_{\rm av}$ obtained with the Grassmann product beamformer codebooks with that of the DFT  KP codebooks~\cite{choi2015advanced} (referred to as `KP-DFT'), and the codebooks generated based on the Grassmannian line packings (GLP) for correlated channel~\cite{love2006limited} (referred to as `Corr-GLP'). For Corr-GLP, the channel correlation matrix ${\bf R}$ is calculated from  ${\cal H}_{\rm train}$ as ${\bf R} = \E_{\bf H}\left({{\bf H}^H{\bf H}}\right)$. It was not possible to show the performance of the Corr-GLP codebooks for large $M_v , M_h$ because finding the GLP in large dimensions is extremely computation intensive. The KP-DFT  codebooks are  simple to construct but is outperfromed by our method. This is because  the KP-DFT  codebooks contain only the beams lying in the direction of the right and left dominant singular vectors of the reshaped FD-MISO channel $\tilde{\bf H}$ as given in~\eqref{eq::prod::codeword}. 

In Fig.~\ref{fig::Results::Rate::Mr2sweep} and~\ref{fig::Results::Rate::Mr3sweep}, we plot the normalized mutual information gain obtained with the product precoder codebooks with varying SNR at different feedback bit allocations and Tx antenna configurations. We observe that the performance of the precoder codebooks approach the gain with unquantized product precoders as the number of feedback bits and SNR increase. The sub-optimality of the product codebooks is caused by the loss in beamforming gain and mutual information by the approximation with the unquantized beamformer (Sec.~\ref{sec::UnquantBeamformer}) and precoder (Lem.~\ref{cor::RateMaximization}). In Fig.~\ref{fig::Results::Rate::Mr2} and~\ref{fig::Results::Rate::Mr3}, we compare the performance of the product codebook and the VQ codebook. 
As expected, $R_{\rm av}$ for the product precoder codebook is slightly worse than $R_{\rm av}$ of the  VQ codebook. This is expected because the VQ works directly on the space of optimal precoders obtained from ${\cal H}_{\rm train}$ while in our method, some accuracy is lost while finding the representation of the product precoder in the TPM.  However, as discussed in detail already in Remark~\ref{Remark::ComplexityAnalysis}, the VQ codebook construction  is significantly more computation intensive than our codebook, as $M_v,M_h$ are large, with diminishing gains in ${R_{\rm av}}$ as seen in Fig.~\ref{fig::Results::Rate::Mr2sweep},~\ref{fig::Results::Rate::Mr3sweep},~\ref{fig::Results::Rate::Mr2},~\ref{fig::Results::Rate::Mr3}. To demonstrate the difference in complexity, in Fig.~\ref{fig::Results::RunTime}, we compare the run-time of the construction of the codebooks using the VQ method~\cite{roh2006design} and the Grassmann product codebooks for different antenna configurations and codebook sizes. The run-times were obtained by averaging the run-times of the codebook construction algorithms over $500$ iterations in the same computation environment. In order to obtain a unit-free measure, we normalized the absolute run-times by  dividing  them  with  the  average absolute  run-time  of the Grassmann product codebook for $M_v \times M_h = 3 \times 3$ with $[B_v,B_h] = [3,3]$. As is evident from this discussion, the VQ method will not scale to large antenna configurations, whereas our method will work well in those cases as well.

\section{Conclusion}
In this paper, we explored the classical problem of precoder codebook design in FDD FD-MIMO systems. Given a dataset of channel realizations, this problem  has been identified as an application of ML in physical layer communication. However, the ``black-box'' application of the ML techniques, such as  DL, may not be beneficial since these techniques tend to work well in Euclidean domain whereas the optimal precoders exist on a GM.  Using the tensor representation of the channel, we showed that the precoder can be approximated as an element in a TPM. This product representation allows us to construct codebooks in the factor manifolds, significantly reducing the complexity compared to the traditional codebook construction methods, such as VQ. We show that finding the codebooks in the factor manifolds is equivalent to $K$-means clustering in the factor GMs with chordal distance metric. This work can be extended in various directions. {\em First}, the codebook can be designed for dual polarized antennas which are more realistic assumptions in cellular systems. {\em Second}, the codebook update methods should be designed such that the codebook adapts to the non-stationary channel distributions. {\em Third}, from the ML perspective, it would be interesting to pose the problem as training an autoencoder. However, following the ideas of theory-guided ML, the challenge will be to constraint the autoencoders to generate the codebooks in a topological manifold, such as GM in this case.

\appendix

\subsection{Proof of Lem.~\ref{lemm::BF::prod::codebook::design}}
\label{app::BF::prod::codebook::design}
From Def.~\ref{def::grassmann::BF::productcodebook}, $\hat{\cal F} = \hat{\cal F}_v \times \hat{\cal F}_h = \underset{{\cal F}_v, {\cal F}_h} \argmin\ L_{\rm ub}({\cal F}) = \underset{{\cal F}_v, {\cal F}_h}\argmin\ \E_{\tilde{\bf u}_1, \tilde{\bf v}_1}\ \left[ \underset{{\substack{{\bf f}_v \in {\cal F}_v \\ {\bf f}_v \in {\cal F}_v}}} \min \left( d^2_c({\tilde{\bf u}}^*_1, {\bf f}_v) + d^2_c({\tilde{\bf v}}_1, {\bf f}_h) \right) \right]$
$= \underset{{\cal F}_v, {\cal F}_h} \argmin\ \E_{\tilde{\bf u}_1}\ \left[ \underset{{\bf f}_v \in {\cal F}_v} \min\ d^2_c({\tilde{\bf u}}^*_1, {\bf f}_v) \right] + \E_{\tilde{\bf v}_1}\ \left[ \underset{{\bf f}_h \in {\cal F}_h} \min\ d^2_c({\tilde{\bf v}}_1, {\bf f}_h) \right]. $
This objective  can be minimized if both the terms in the summation are independently minimized. Therefore the codebooks $\hat{\cal F}_v, \hat{\cal F}_h$ that form the Grassmann product codebook $\hat{\cal F}$ are given as \begin{align}
     \hat{\cal F}_v &= \underset{{\substack{{\cal F}_v \subseteq {\cal G}(M_v, 1) \\ |{\cal F}_v| = 2^{B_v}}}}\argmin \E_{\tilde{\bf u}_1}\ \left[ \underset{{\bf f}_v \in {\cal F}_v} \min\ d^2_c({\tilde{\bf u}}^*_1, {\bf f}_v) \right], \quad \hat{\cal F}_h = \underset{{\substack{{\cal F}_h \subseteq {\cal G}(M_h, 1) \\ |{\cal F}_h| = 2^{B_h}}}}\argmin \E_{\tilde{\bf v}_1}\ \left[ \underset{{\bf f}_h \in {\cal F}_h} \min\ d^2_c({\tilde{\bf v}}_1, {\bf f}_h) \right].
\end{align}
Comparing the general Grassmannian $K$-means objective in \eqref{eq::Kmeans::objective} in Sec.~\ref{sec::Kmeans} and the above codebook design criteria, $\hat{\cal F}_h, \hat{\cal F}_v$ can be found by the $K$-means clustering algorithm for $[K, n, k] = [2^{B_h}, M_h, 1], [2^{B_v}, M_v, 1]$ respectively, in Alg.~\ref{alg::Product::KmeansAlgo}. Therefore we have $\hat{\cal F}_h = {\cal F}^K_h, \hat{\cal F}_v = {\cal F}^K_v, \text{ and } \hat{\cal F} = \hat{\cal F}_v \times \hat{\cal F}_h = {\cal F}^K_v \times {\cal F}^K_h$ and the criteria for the choosing the optimal beamformer $\hat{\bf f}$ from $\hat{\cal F}_v$, $\hat{\cal F}_h$ for a given ${\bf H}$ as $\hat{\bf f}_v = \underset{{\bf f} \in \hat{\cal F}_v} \argmin\ d^2_c({\tilde{\bf u}}^*, {\bf f}), \hat{\bf f}_h = \underset{{\bf f} \in \hat{\cal F}_h} \argmin\ d^2_c({\tilde{\bf v}}, {\bf f}), \hat{\bf f} = \hat{\bf f}_v \otimes \hat{\bf f}_h$.

\subsection{Proof of Lem.~\ref{lemm::PC::prod::codebook::design}}
\label{app::pc::prod::codebook::design}
From Def.~\ref{def::grassmann::PC::productcodebook} and \eqref{eq::DesignCriterion::GrassmannEq}, we modify the optimization objective according to the chordal distance approximation in Assum.~\ref{rem::ChordalDist::Approx} which gives the following codebook design criterion.
\begin{align}
  \hat{\cal F}_{v} \times \hat{\cal F}_{h} &=   \underset{{\cal F} \subseteq {\cal G}^{\times}\left((M_v,M_h), (\texttt{r},\texttt{r})\right)} \argmin\ \underset{Q(\cdot)} \min \ \E_{{\bf H}}\left[d^2_c\left(\left(\bar{\bf A}^{(2)}, \bar{\bf A}^{(1)}\right), \left(Q(\bar{\bf A}^{(2)}), Q(\bar{\bf A}^{(1)})\right)\right)\right] \label{eq::Kmeans::simplification} \\
    &=  \underset{{\substack{{\cal F}_{h} \subseteq {\cal G}(M_h, \texttt{r}) \\ {\cal F}_{v} \subseteq {\cal G}(M_v, \texttt{r})}}} \argmin\ \underset{Q(\cdot)} \min \ \E_{\bar{\bf A}^{(2)}}\left[d^2_c\left(\bar{\bf A}^{(2)}, Q(\bar{\bf A}^{(2)})\right)\right] + \E_{\bar{\bf A}^{(1)}}\left[d^2_c\left(\bar{\bf A}^{(1)}, Q(\bar{\bf A}^{(1)})\right) \right]. \notag 
\end{align}
Thus the design criteria for $\hat{\cal F}_{h}$, $\hat{\cal F}_{v}$  for ${\bar{\bf A}}^{(1)}$, ${\bar{\bf A}}^{(2)}$ is 
\begin{align*}
    \hat{\cal F}_{h} = \underset{{\substack{{\cal F}_{h} \subseteq {\cal G}(M_h, \texttt{r}) \\ |{\cal F}_{h}| = 2^{B_h}}}} \argmin\ \underset{Q(\cdot)} \min \ \E_{\bar{\bf A}^{(1)}}\left[d^2_c\left(\bar{\bf A}^{(1)}, Q(\bar{\bf A}^{(1)})\right)\right],\ \hat{\cal F}_{v} = \underset{{\substack{{\cal F}_{v} \subseteq {\cal G}(M_v, \texttt{r}) \\ |{\cal F}_{v}| = 2^{B_v}}}} \argmin\ \underset{Q(\cdot)} \min \ \E_{\bar{\bf A}^{(2)}}\left[d^2_c\left(\bar{\bf A}^{(2)}, Q(\bar{\bf A}^{(2)})\right) \right].\label{eq::CB::DesignCriterion::dist} 
\end{align*}
Comparing the general Grassmannian $K$-means clustering objective in~\eqref{eq::Kmeans::objective} in Sec.~\ref{sec::Kmeans} with the above codebook design criteria for $\hat{\cal F}_{h}$, $\hat{\cal F}_{v}$, we have $\hat{\cal F}_{h} = {\cal F}^K_{h} {\text{ for }} [K, n, k] = [2^{B_h}, M_h, \texttt{r}], \hat{\cal F}_{v} = {\cal F}^K_{v} {\text{ for }} [K, n, k] = [2^{B_v}, M_v, \texttt{r}]$, thus $\hat{\cal F} = \hat{\cal F}_{v} \times \hat{\cal F}_{h} = {\cal F}^K_{v} \times {\cal F}^K_{h}$ and the corresponding optimal quantizers for $\bar{\bf A}^{(1)}$, $\bar{\bf A}^{(2)}$ that minimize the average distortion are  $Q(\bar{\bf A}^{(1)}) = \underset{{\bf F} \in \hat{\cal F}_{h}} \argmin\ d^2_c(\bar{\bf A}^{(1)}, {\bf F})$, $Q(\bar{\bf A}^{(2)}) = \underset{{\bf F} \in \hat{\cal F}_{v}} \argmin\ d^2_c(\bar{\bf A}^{(2)}, {\bf F})$.

\section*{Acknowledgment}
The authors would like to thank Andreas F. Molisch for his valuable comments on the conference version of this paper.

\setstretch{1.25}
\bibliographystyle{IEEEtran}  
\bibliography{Journal-vTComm}

\end{document}

%% file: notationITA.tex
\let\emptyset\varnothing
\newcommand\norm[1]{\left\lVert#1\right\rVert}

\def\nb0{{\mathbf{0}}}
\def\nb1{{\mathbf{1}}}

\newtheorem{lemma}{Lemma}

\newtheorem{ndef}{Definition}

\newtheorem{prop}{Proposition}

\newtheorem{remark}{Remark}
\newtheorem{assumption}{Assumption}

\def\argmin{\operatorname{arg~min}}
\def\argmax{\operatorname{arg~max}}
\def\E{\mathbb{E}}

\def\R{\mathbb{R}}

\def\orthvect{{\cal U}(M_t,1)}

\def\C{{\mathbb C}}

\def\C{{\mathbb C}}
\def\R{{\mathbb R}}
\def\nscrX{{\mathscr{X}}}
\def\nscrY{{\mathscr{Y}}}
\def\nscrG{{\mathscr{G}}}
\def\nscrH{{\mathscr{H}}}

%% file: Journal-vTComm.bbl
\begin{thebibliography}{10}
\providecommand{\url}[1]{#1}
\csname url@samestyle\endcsname
\providecommand{\newblock}{\relax}
\providecommand{\bibinfo}[2]{#2}
\providecommand{\BIBentrySTDinterwordspacing}{\spaceskip=0pt\relax}
\providecommand{\BIBentryALTinterwordstretchfactor}{4}
\providecommand{\BIBentryALTinterwordspacing}{\spaceskip=\fontdimen2\font plus
\BIBentryALTinterwordstretchfactor\fontdimen3\font minus
  \fontdimen4\font\relax}
\providecommand{\BIBforeignlanguage}[2]{{%
\expandafter\ifx\csname l@#1\endcsname\relax
\typeout{** WARNING: IEEEtran.bst: No hyphenation pattern has been}%
\typeout{** loaded for the language `#1'. Using the pattern for}%
\typeout{** the default language instead.}%
\else
\language=\csname l@#1\endcsname
\fi
#2}}
\providecommand{\BIBdecl}{\relax}
\BIBdecl

\bibitem{bhogi2020learning}
K.~Bhogi, C.~Saha, and H.~S. Dhillon, ``Learning on a {G}rassmann manifold:
  {CSI} quantization for massive {MIMO} systems,'' in \emph{Proc. 54th Asilomar
  Conf. on Signals, Systems, and Computers}, 2020, pp. 179--186.

\bibitem{dorner2017deep}
S.~D{\"o}rner, S.~Cammerer, J.~Hoydis, and S.~Ten~Brink, ``Deep learning based
  communication over the air,'' \emph{IEEE J. of Sel. Topics in Signal
  Process.}, vol.~12, no.~1, pp. 132--143, 2017.

\bibitem{o2017introduction}
T.~O`Shea and J.~Hoydis, ``An introduction to deep learning for the physical
  layer,'' \emph{IEEE Trans. on Cognitive Commun. and Networking}, vol.~3,
  no.~4, pp. 563--575, 2017.

\bibitem{karpatne2017theory}
A.~Karpatne \emph{et~al.}, ``Theory-guided data science: A new paradigm for
  scientific discovery from data,'' \emph{IEEE Trans. on Knowledge and Data
  Engineering}, vol.~29, no.~10, pp. 2318--2331, 2017.

\bibitem{narula1998efficient}
A.~Narula, M.~J. Lopez, M.~D. Trott, and G.~W. Wornell, ``Efficient use of side
  information in multiple-antenna data transmission over fading channels,''
  \emph{IEEE J. on Sel. Areas in Commun.}, vol.~16, no.~8, pp. 1423--1436, Oct
  1998.

\bibitem{love2008overview}
D.~J. {Love}, R.~W. {Heath}, V.~K. {N. Lau}, D.~{Gesbert}, B.~D. {Rao}, and
  M.~{Andrews}, ``An overview of limited feedback in wireless communication
  systems,'' \emph{IEEE J. on Sel. Areas in Commun.}, vol.~26, no.~8, pp.
  1341--1365, Oct 2008.

\bibitem{deepCSI2018}
C.~Wen, W.~Shih, and S.~Jin, ``Deep learning for massive {MIMO} {CSI}
  feedback,'' \emph{IEEE Wireless Commun. Letters}, vol.~7, no.~5, pp.
  748--751, Oct 2018.

\bibitem{deepCSI2019}
T.~Wang, C.~Wen, S.~Jin, and G.~Y. Li, ``Deep learning-based {CSI} feedback
  approach for time-varying massive {MIMO} channels,'' \emph{IEEE Wireless
  Commun. Letters}, vol.~8, no.~2, pp. 416--419, Apr 2019.

\bibitem{goodfellow2016deep}
I.~Goodfellow, Y.~Bengio, A.~Courville, and Y.~Bengio, \emph{Deep
  learning}.\hskip 1em plus 0.5em minus 0.4em\relax MIT press Cambridge, 2016,
  vol.~1, no.~2.

\bibitem{love2005limited}
D.~J. Love and R.~W. Heath, ``Limited feedback unitary precoding for spatial
  multiplexing systems,'' \emph{IEEE Trans. on Inf. Theory}, vol.~51, no.~8,
  pp. 2967--2976, Aug 2005.

\bibitem{love2003grassmannian}
D.~J. {Love}, R.~W. {Heath}, and T.~{Strohmer}, ``Grassmannian beamforming for
  multiple-input multiple-output wireless systems,'' \emph{IEEE Trans. on Inf.
  Theory}, vol.~49, no.~10, pp. 2735--2747, Oct 2003.

\bibitem{love2006limited}
D.~J. Love and R.~W. Heath, ``Limited feedback diversity techniques for
  correlated channels,'' \emph{IEEE Trans. on Veh. Tech.}, vol.~55, no.~2, pp.
  718--722, Mar 2006.

\bibitem{amiri2008adaptive}
K.~Amiri, D.~Shamsi, B.~Aazhang, and J.~R. Cavallaro, ``Adaptive codebook for
  beamforming in limited feedback {MIMO} systems,'' in \emph{Proc. 42nd Annual
  Conf. on Inf. Sciences and Systems}, 2008, pp. 994--998.

\bibitem{mcnamara2002spatial}
D.~P. McNamara, M.~A. Beach, and P.~N. Fletcher, ``Spatial correlation in
  indoor {MIMO} channels,'' in \emph{Proc. IEEE PIMRC}, vol.~1, 2002, pp.
  290--294.

\bibitem{3gpp.25.996}
\emph{Spatial channel model for {M}ultiple {I}nput {M}ultiple {O}utput ({MIMO})
  simulations}, 3GPP TR 25.996, 2003.

\bibitem{shuang2011design}
T.~Shuang, T.~Koivisto, H.~L. Maattanen, K.~Pietikainen, T.~Roman, and
  M.~Enescu, ``Design and evaluation of {LTE}-{A}dvanced double codebook,'' in
  \emph{IEEE 73rd Veh. Technol. Conf.}, 2011, pp. 1--5.

\bibitem{ying2014kronecker}
D.~{Ying}, F.~W. {Vook}, T.~A. {Thomas}, D.~J. {Love}, and A.~{Ghosh},
  ``Kronecker product correlation model and limited feedback codebook design in
  a 3{D} channel model,'' in \emph{Proc. IEEE ICC}, Aug 2014, pp. 5865--5870.

\bibitem{li2013codebook}
J.~Li, X.~Su, J.~Zeng, Y.~Zhao, S.~Yu, L.~Xiao, and X.~Xu, ``Codebook design
  for uniform rectangular arrays of massive antennas,'' in \emph{Proc. IEEE
  77th Veh. Technol. Conf.}, 2013, pp. 1--5.

\bibitem{su2013limited}
X.~Su, J.~Zeng, J.~Li, L.~Rong, L.~Liu, X.~Xu, and J.~Wang, ``Limited feedback
  precoding for massive {MIMO},'' \emph{Int. J. of Antennas and Propagation},
  vol. 2013, Oct 2013.

\bibitem{song2018advanced}
J.~Song, J.~Choi, T.~Kim, and D.~J. Love, ``Advanced quantizer designs for
  {FDD}-based {FD-MIMO} systems using uniform planar arrays,'' \emph{IEEE
  Trans. on Signal Process.}, vol.~66, no.~14, pp. 3891--3905, 2018.

\bibitem{choi2015advanced}
J.~{Choi}, K.~{Lee}, D.~J. {Love}, T.~{Kim}, and R.~W. {Heath}, ``Advanced
  limited feedback designs for {FD-MIMO} using uniform planar arrays,'' in
  \emph{Proc. IEEE {GLOBECOM}}, 2015, pp. 1--6.

\bibitem{roh2006design}
J.~C. Roh and B.~D. Rao, ``Design and analysis of {MIMO} spatial multiplexing
  systems with quantized feedback,'' \emph{IEEE Trans. on Signal Process.},
  vol.~54, no.~8, pp. 2874--2886, 2006.

\bibitem{dhillon2003diametrical}
I.~S. Dhillon, E.~M. Marcotte, and U.~Roshan, ``Diametrical clustering for
  identifying anti-correlated gene clusters,'' \emph{Bioinformatics}, vol.~19,
  no.~13, pp. 1612--1619, 2003.

\bibitem{bellman2015adaptive}
R.~E. Bellman, \emph{Adaptive control processes: a guided tour}.\hskip 1em plus
  0.5em minus 0.4em\relax Princeton University Press, 2015, vol. 2045.

\bibitem{yang2019deepCMC}
Q.~Yang, M.~B. Mashhadi, and D.~G{\"u}nd{\"u}z, ``Deep convolutional
  compression for massive {MIMO} {CSI} feedback,'' in \emph{Proc. 29th IEEE
  Int. Workshop on Machine Learning for Signal Process. (MLSP)}, 2019, pp.
  1--6.

\bibitem{guo2020convolutional}
J.~Guo, C.~K. Wen, S.~Jin, and G.~Y. Li, ``Convolutional neural network-based
  multiple-rate compressive sensing for massive {MIMO} {CSI} feedback: Design,
  simulation, and analysis,'' \emph{IEEE Trans. on Wireless Commun.}, vol.~19,
  no.~4, pp. 2827--2840, 2020.

\bibitem{araujo2019tensor}
D.~C. Ara{\'u}jo, A.~L. De~Almeida, J.~P. Da~Costa, and R.~T. de~Sousa,
  ``Tensor-based channel estimation for massive {MIMO-OFDM} systems,''
  \emph{IEEE Access}, vol.~7, pp. 42\,133--42\,147, 2019.

\bibitem{de2003blind}
A.~de~Baynast, L.~De~Lathauwer, and B.~Aazhang, ``Blind {PARAFAC} receivers for
  multiple access-multiple antenna systems,'' in \emph{2003 IEEE 58th Veh.
  Technol. Conf.}, vol.~2, 2003, pp. 1128--1132.

\bibitem{STMCAlmeida}
A.~L.~F. {de Almeida}, G.~{Favier}, and J.~C.~M. {Mota}, ``Space-time
  multiplexing codes: A tensor modeling approach,'' in \emph{2006 IEEE 7th
  Workshop on Signal Process. Advances in Wireless Commun.}, 2006, pp. 1--5.

\bibitem{TD2015Cichoki}
A.~Cichocki, D.~Mandic, L.~D. Lathauwer, G.~Zhou, Q.~Zhao, C.~Caiafa, and H.~A.
  Phan, ``Tensor decompositions for signal processing applications: From
  two-way to multiway component analysis,'' \emph{IEEE Signal Process. Mag.},
  vol.~32, no.~2, pp. 145--163, 2015.

\bibitem{sidiropoulos2017tensor}
N.~D. Sidiropoulos, L.~De~Lathauwer, X.~Fu, K.~Huang, E.~E. Papalexakis, and
  C.~Faloutsos, ``Tensor decomposition for signal processing and machine
  learning,'' \emph{IEEE Trans. on Signal Process.}, vol.~65, no.~13, pp.
  3551--3582, 2017.

\bibitem{lui2012human}
Y.~M. Lui, ``Human gesture recognition on product manifolds,'' \emph{JMLR},
  vol.~13, no.~1, pp. 3297--3321, 2012.

\bibitem{tse2000performance}
C.~H. Tse, K.~W. Yip, and T.~S. Ng, ``Performance tradeoffs between maximum
  ratio transmission and switched-transmit diversity,'' in \emph{Proc. IEEE
  PIMRC}, vol.~2, 2000, pp. 1485--1489.

\bibitem{andersen2000antenna}
J.~B. {Andersen}, ``Antenna arrays in mobile communications: gain, diversity,
  and channel capacity,'' \emph{IEEE Antennas and Propag. Mag.}, vol.~42,
  no.~2, pp. 12--16, Apr 2000.

\bibitem{simon2005digital}
M.~K. Simon and M.~S. Alouini, \emph{Digital communication over fading
  channels}.\hskip 1em plus 0.5em minus 0.4em\relax John Wiley \& Sons, 2005,
  vol.~95.

\bibitem{cover1999elements}
T.~M. Cover, \emph{Elements of Information Theory}.\hskip 1em plus 0.5em minus
  0.4em\relax John Wiley \& Sons, 1999.

\bibitem{scaglione2002optimal}
A.~Scaglione, P.~Stoica, S.~Barbarossa, G.~B. Giannakis, and H.~Sampath,
  ``Optimal designs for space-time linear precoders and decoders,'' \emph{IEEE
  Trans. on Signal Process.}, vol.~50, no.~5, pp. 1051--1064, 2002.

\bibitem{kapetanovic2010comparison}
D.~Kapetanovic and F.~Rusek, ``A comparison between unitary and non-unitary
  precoder design for {MIMO} channels with {MMSE} detection and limited
  feedback,'' in \emph{Proc. IEEE GLOBECOM}, 2010, pp. 1--6.

\bibitem{kapteyn1986approach}
A.~Kapteyn, H.~Neudecker, and T.~Wansbeek, ``An approach ton-mode components
  analysis,'' \emph{Psychometrika}, vol.~51, no.~2, pp. 269--275, 1986.

\bibitem{eckart1936approximation}
C.~Eckart and G.~Young, ``The approximation of one matrix by another of lower
  rank,'' \emph{Psychometrika}, vol.~1, no.~3, pp. 211--218, 1936.

\bibitem{kolda2009tensor}
T.~G. Kolda and B.~W. Bader, ``Tensor decompositions and applications,''
  \emph{SIAM review}, vol.~51, no.~3, pp. 455--500, 2009.

\bibitem{edelman1998geometry}
A.~Edelman, T.~A. Arias, and S.~T. Smith, ``The geometry of algorithms with
  orthogonality constraints,'' \emph{SIAM J. on Matrix Anal. and Appl.},
  vol.~20, no.~2, pp. 303--353, 1998.

\bibitem{LBGAlgorithm}
Y.~Linde, A.~Buzo, and R.~Gray, ``An algorithm for vector quantizer design,''
  \emph{IEEE Trans. on Commun.}, vol.~28, no.~1, pp. 84--95, Jan 1980.

\bibitem{mondal2007quantization}
B.~{Mondal}, S.~{Dutta}, and R.~W. {Heath}, ``Quantization on the {G}rassmann
  manifold,'' \emph{IEEE Trans. on Signal Process.}, vol.~55, no.~8, pp.
  4208--4216, Aug 2007.

\bibitem{nemhauser1978best}
G.~L. Nemhauser and L.~A. Wolsey, ``Best algorithms for approximating the
  maximum of a submodular set function,'' \emph{Mathematics of Operations
  Research}, vol.~3, no.~3, pp. 177--188, 1978.

\bibitem{nemhauser1978analysis}
G.~L. Nemhauser, L.~A. Wolsey, and M.~L. Fisher, ``An analysis of
  approximations for maximizing submodular set functions--{I},''
  \emph{Mathematical Programming}, vol.~14, no.~1, pp. 265--294, 1978.

\bibitem{roh2006transmit}
J.~C. Roh and B.~D. Rao, ``Transmit beamforming in multiple-antenna systems
  with finite rate feedback: A {VQ}-based approach,'' \emph{IEEE Trans. on Inf.
  Theory}, vol.~52, no.~3, pp. 1101--1112, 2006.

\bibitem{konar2017greed}
A.~Konar and N.~D. Sidiropoulos, ``Greed is good: Leveraging submodularity for
  antenna selection in massive {MIMO},'' in \emph{Proc. 51st Asilomar Conf. on
  Signals, Systems, and Computers}, 2017, pp. 1522--1526.

\bibitem{curtef2012riemannian}
O.~Curtef, G.~Dirr, and U.~Helmke, ``Riemannian optimization on tensor products
  of {G}rassmann manifolds: {A}pplications to generalized
  {R}ayleigh--quotients,'' \emph{SIAM J. on Matrix Anal. and Appl.}, vol.~33,
  no.~1, pp. 210--234, 2012.

\bibitem{alkhateeb2019deepmimo}
A.~Alkhateeb, ``Deep{MIMO}: {A} generic deep learning dataset for millimeter
  wave and massive {MIMO} applications,'' in \emph{Proc. ITA}, Feb 2019, pp.
  1--8.

\end{thebibliography}
